\documentclass[fleqn,usenatbib]{mnras}

\usepackage[T1]{fontenc}
\usepackage{ae,aecompl}

\usepackage{graphicx}	
\usepackage{amsmath}	
\usepackage{amssymb}	
\usepackage{bm}
\usepackage{times}
\usepackage{threeparttable}
\usepackage{multicol} 
\usepackage{ulem}
\usepackage{booktabs}
\usepackage{comment}
\usepackage{multirow}
\usepackage{hyperref} 
\usepackage[title]{appendix}

\def\nus{\nu_{\rm s}}
\def\nuic{\nu_{_{\rm IC}}}

\def\gc{\gamma_{\rm c}}
\def\gm{\gamma_{\rm m}}

\def\Niso{N_{e,\rm iso}}
\def\gc{\gamma_{\rm c}}
\def\hgc{\widehat{\gamma_{\rm c}}}

\def\gm{\gamma_{\rm m}}
\def\hgm{\widehat{\gamma_{\rm m}}}
\def\hnm{\widehat{\nu_{\rm m}}}
\def\g0{\gamma_{0}}
\def\reps{\frac{\epsilon_e}{\epsilon_B}}
\def\mrow{\multirow}
\def\mcol{\multicolumn}
\def\cent{\centering}

\usepackage{newtxtext,newtxmath}
\usepackage{amsmath}
\usepackage{breqn}
\usepackage{extarrows}
\usepackage{amssymb}	
\usepackage{mathrsfs}
\usepackage{color}

\def\beq{\begin{equation}}
\def\eeq{\end{equation}}
\def\beqn{\begin{eqnarray}}
\def\eeqn{\end{eqnarray}}

\title[Analytic SSC modeling]{Analytic Modeling of Synchrotron-Self-Compton Spectra: Application to GRB 190114C}

\author[Yamasaki \& Piran]
{Shotaro Yamasaki\thanks{E-mail: shotaro.yamasaki@mail.huji.ac.il}$^{1}$ and Tsvi Piran$^{1}$
\\
$^{1}$Racah Institute of Physics, The Hebrew University of Jerusalem, Jerusalem 91904, Israel\\
}

\date{Accepted XXX. Received YYY; in original form ZZZ}

\pubyear{2020}

\begin{document}
\label{firstpage}
\pagerange{\pageref{firstpage}--\pageref{lastpage}}
\maketitle

\begin{abstract}
Observations of TeV emission from early gamma-ray burst (GRB) afterglows revealed the long sought for  inverse Compton (IC) upscattering of the lower energy synchrotron. 
However, it turned out that {the long hoped for ability to easily interpret}
the synchrotron-self-Compton  (SSC) spectra {didn't materialize}.
The TeV emission is in the Klein-Nishina (KN) regime {and}
the simple Thomson regime SSC spectrum, 
{is modified}, 
complicating the scene. We describe here a methodology to obtain an analytic approximation to an observed spectrum and infer the conditions at the emitting region. The methodology is general and can be used in  any such source. As a test case we apply it to the observations of GRB 190114C. 
We find that the procedure of fitting the model parameters using the  analytic SSC spectrum suffers from some generic problems.
However, at the same time, it  conveniently gives a useful insight into the conditions that shape the spectrum.  Once we introduce a correction to the standard KN approximation, the best fit solution is consistent with the one found in detailed numerical simulations.  As in the numerical analysis, we find a family of solutions that provide a good approximation to the data and  satisfy roughly $B\propto \Gamma^{-3}$ between the magnetic field and the bulk Lorentz factor, and we provide a tentative explanation why such a family arises.  

\end{abstract}

\begin{keywords}
radiation mechanisms: non-thermal-- gamma-ray burst: individual: GRB 190114C
\end{keywords}


\section{Introduction}

Recent observations by Cherenkov detectors
\citep{magic19,hess21}   of TeV gamma-rays from several Gamma-Ray Bursts (GRBs) revealed that  synchrotron-self Compton (SSC) emission  plays an important role in shaping the early stages of GRB afterglows. 
A full analysis of the spectrum that takes care of the inverse Compton (IC) cooling,  Klein-Nishina (KN) correction to the cross section, self absorption and emission of the created pairs required detailed numerical analysis.   However,  key information about the characteristic  frequencies, which is crucial for the qualitative understanding of the system, is often smeared out and lost in the smooth spectrum resulting from the numerical analysis. 
In order to determine those we 
apply here the analytic understanding of SSC modeling that includes the effects of KN cross-section on the electron distribution as described by \cite{nakar09}, hereafter denoted NAS09. 

At first sight it looks as if the combination of synchrotron and SSC will fully constrain the conditions within the emitting region from a single observations of the SED. The IC component provides at least two additional conditions (the  peak flux and its frequency) and together with the synchrotron observables these should enable us to solve for the  parameters such as the magnetic field and the energy density of the relativistic electrons uniquely. Indeed,  the preliminary GRB 190114C data  with a low energy peak around 10 keV and a high energy one around 1 TeV, suggested that the typical Lorentz factor (LF) of the emitting electrons is  $\sim 10^4$ and the bulk flow LF is $\sim 100$ \citep{dp19}, which put the IC component at the edge of the KN regime. In such a case, the  KN corrections to IC scatterings, and in particular the feedback of the IC in the KN regime on the electron's distribution,  must be taken into account  (\citealt{Derishev2001,Derishev2003}; NAS09). As we see below, however, this significantly complicates the interpretation of the spectrum.  
While these KN effects are generic and they exist also in numerical solutions, they are particularly cumbersome in the analytic approach.

In spite of these difficulties we describe here a methodology for determining  the physical conditions within the emitting region using
analytic solutions to the SSC problem  based on the analytic  model of NAS09. 
We slightly modify the original NAS09 model introducing a revision to common approximation concerning  KN effects. This takes into account the fact that KN effect begins at energies lower than the electron's rest mass energy. We apply this model to GRB 190114C and discuss the inferred physical parameters of the afterglow's emitting region. We note that while our discussion focuses on GRB afterglows in principle specific features of GRB afterglows are used only in one place and that the general ideas can be applied to other systems in which SSC in the KN regime arises. 

The paper is organized as follows. In \S \ref{s:model}, we describe briefly the NAS09 model and the methodology.
We find analytic fits for the   GRB 190114C SED in \S \ref{s:GRB190114C} and compare the analytic results to numerical ones in \S \ref{s:comparison}. 
We then discuss and summarize our findings in \S \ref{s:discussion}.

\section{The Analytic SSC Model}
\label{s:model}

\subsection{Overview of NAS09}
 
\label{ss:model}

\citet{sari01} considered SSC modeling within the context of GRBs for cases in which KN effects are unimportant. This was generalized by NAS09  \cite[see also][]{Derishev2001,Derishev2003} who obtained an analytic approximation of synchrotron and SSC spectra including KN effects. We begin by reviewing essential features of this model on which this work is based. It is assumed that the electrons are populated  by a steady injection of relativistic
electrons with a power-law distribution with $dN_e/d\gamma\propto \gamma^{-p}$ ($p>2$) above the minimal Lorentz factor (LF)  of electrons $\gm$.
The observed synchrotron frequency of  an electron with LF $\gamma$ is: 
\beqn
\nu_{\rm syn}(\gamma)\equiv
\frac{ \Gamma\gamma^2 q_e B }{2\pi m_e c (1+z)}\ ,
\eeqn
{where $\Gamma$ is the bulk LF of the source, $B$ is the magnetic field, and $z$ is the redshift of the source (with electron mass $m_e$ and speed of light $c$).}
{In the presence of the KN feedback, the SSC to the synchrotron emissivity ratio  (the Compton Y parameter\footnote{Note that this definition of the Compton Y parameter is different from the common definition, the mean number of scatters suffered by a photon before escape times the net addition of energy per scatter. However, in our case the optical depth is less than unity and because of KN effects, namely when a photon has energy $\gg m_e c^2$ the Klein-Nishina cross section for an encounter with a highly relativistic electron becomes extremely small, and a second scattering is always suppressed. In this case our definition approximates well the regular one. })  depends on $\gamma$:
\beqn
\label{eq:Y}
Y(\gamma)\equiv \frac{P_{\rm SSC}(\gamma)}{P_{\rm syn}(\gamma)} \  .
\eeqn
The cooling LF, $\gc$ (calculated taking also IC cooling into account), is defined such that electrons above $\gc$ cool efficiently over the system lifetime:
\beqn
\label{eq:gc}
\gc\left[1+Y(\gc)\right] = \gamma_{\rm c}^{\rm syn} = \frac{6\pi m_ec(1+z)}{\sigma_{\rm T}t_{\rm obs}\Gamma B^2},
\eeqn
where $\gamma_{\rm c}^{\rm syn}$ is the cooling LF ignoring IC cooling, $t_{\rm obs}$ is the observing time  and $\sigma_{\rm T}$ the Thomson cross section.

Unlike the rather simple synchrotron spectra or even SSC with no KN corrections, the SSC spectra in the KN regime is quite complicated.  It takes different shapes depending on the interplay between the different characteristic electron's LF and the corresponding frequencies.  In addition to the common typical and cooling electron LFs, $\gm$, and $\gc$ and their corresponding frequencies we have several other characteristic LFs. In particular following NAS09 we define $\widehat{\gamma}$,
the maximal LF  of an electron that can efficiently upscatter a synchrotron photon emitted by an electron with LF $\gamma$: 
\beqn
\label{eq:hat_gamma}
\widehat{\gamma}(\gamma)\equiv\frac{\Gamma f_{_{\rm KN}} \,m_e c^2}{h\nu_{\rm syn}(\gamma){\,(1+z)}}\propto\frac{1}{B\gamma^{2}}\  ,
\eeqn
where $h$ is the Planck constant. The factor $f_{_{\rm KN}}<1$ introduced here is a correction factor for the KN limit that will be discussed in \S \ref{ss:f_KN}. 

The KN suppression leads to additional characteristic LFs: $\hgm\equiv\widehat{\gamma}(\gm)$ and $\hgc\equiv\widehat{\gamma}(\gc)$ that are related to the typical and the cooling LFs, respectively. 
We also define  $\gamma_{\rm self}$ as the critical electron LF that satisfies $\widehat{\gamma}(\gamma_{\rm self})=\gamma_{\rm self}$:
\beqn
\label{eq:gamma_self}
 \gamma_{\rm self}=f_{_{\rm KN}}^{1/3}\left(\frac{B_{\rm cr}}{B}\right)^{1/3} \ ,
\eeqn
where $B_{\rm cr}\equiv2\pi m_e^2c^3/(h q_e)\sim4.4\times10^{13}$ G is the Schwinger magnetic field strength. Note that 
$\gamma^{2/3}\widehat{\gamma}^{1/3}= \gamma_{\rm self}$, for any $\gamma $.

Finally, in the presence of the KN feedback $Y(\gamma)$  decreases as $\gamma$ increases.  At a critical LF $\gamma_0$
\beqn
\label{eq:gamma_0}
Y(\gamma_0)=1 \ .
\eeqn
Above $\gamma_0$ SSC cooling becomes ineffective. Note that $\gamma_0$ is well defined only when $Y(\gc)>1$ (in which case $\gc<\gamma_0$). Otherwise SSC cooling is unimportant and all the electrons cool just  by synchrotron emission, namely in this case $\gc=\gamma_{\rm c}^{\rm syn}$.  

The overall spectrum is piecewise linear in the log-log plane with  breaks at the transition frequencies. 
The set of characteristic LFs $\gc$, $\gm$, $\gamma_{\rm self}$, $\gamma_0$, $\hgm$, and $\hgc$ defines the spectral type (NAS09), which are broadly separated into fast ($\gc<\gm$) and slow ($\gm<\gc$) cooling regimes. These are further classified into different spectral cases depending on the significance of the KN effect, i.e., via the ratio $\gm/\hgm = (\gm/\gamma_{\rm self})^3$ for fast cooling and $\gc/\hgc= (\gc/\gamma_{\rm self})^3$ for  slow cooling. The strength of KN effects is quantified by the deviation of the typical (cooling) electron LF $\gm$ ($\gc$) from $\gamma_{\rm self}$. 

\begin{table*}
\centering
\caption{Summary of spectral classification based on NAS09 for $\epsilon_e/\epsilon_B > 1$ (otherwise IC is negligible and we have mostly synchrotron emission). Top panel: fast cooling  ($\gc < \gm$) and bottom panel: slow-cooling ($\gm < \gc$) cooling. }
\label{tab:spectral regime}
\begin{tabular}{@{}|c|c|c|c|c|c|@{}}
\mcol{1}{c}{} & \mcol{1}{c}{\cent  weak} &  \mcol{1}{c}{\cent  $\xlongleftarrow[]{\hspace{6.5em}}$} &  \mcol{1}{c}{\cent  KN effect}  & \mcol{1}{c}{\cent  $\xlongrightarrow[]{\hspace{6.5em}}$} & \mcol{1}{c}{\cent  strong}\vspace{2mm}\\
\hline\hline
\mrow{2}{23mm}{\cent Spectral regime (fast cooling)} & \mrow{2}{14mm}{\cent fast case I (FI)} & \mrow{2}{18mm}{\cent fast case III (FIII)} & \mrow{2}{18mm}{\cent fast case IIc (FIIc)} & \mrow{2}{18mm}{\cent fast case IIb (FIIb)} & \mrow{2}{18mm}{\cent fast case IIa (FIIa)}\\
&  &  &  &  & \\
\hline
\mrow{2}{20mm}{\cent Condition } & \mrow{2}{30mm}{\cent $\frac{\gm}{\hgm}=\gm^3 \left(\frac{B}{f_{_{\rm KN}} B_{\rm cr}}\right)<1$}  & \mrow{2}{25mm}{\cent $1\le\frac{\gm}{\hgm} < \left(\reps\right)^{\frac{1}{3}}$} & \mrow{2}{27mm}{\cent $\left(\reps\right)^{\frac{1}{3}}<\frac{\gm}{\hgm}<\reps$} & \mrow{2}{25mm}{\cent $\reps<\frac{\gm}{\hgm}<\left(\reps\right)^{3}$}  & \mrow{2}{25mm}{\cent $\left(\reps\right)^3<\frac{\gm}{\hgm}$}  \\
&  &  &  &  & \\
\hline
\mrow{2}{20mm}{\cent Frequencies}& \mrow{2}{25mm}{\cent $\gc<\gm<\gamma_{\rm self}<\hgm<\g0$} &\mrow{2}{25mm}{\cent $\gc<\gm\sim\gamma_{\rm self}\sim\hgm<\g0$} &  \mrow{2}{25mm}{\cent $\gc<\hgm<\gamma_{\rm self}<\gm<\g0<\widehat{\hgm}$} &  \mrow{2}{25mm}{\cent $\gc<\hgm<\gamma_{\rm self}<\g0<\gm$} & \mrow{2}{25mm}{\cent $\gc<\hgm<\g0<\gamma_{\rm self}<\gm$} \\
&  &  &  &  & \\
\hline\hline
\mcol{1}{c}{}&\mcol{1}{c}{}&\mcol{1}{c}{}&\mcol{1}{c}{}&\mcol{1}{c}{}&\mcol{1}{c}{}\\
\hline\hline
\mrow{2}{23mm}{\cent Spectral regime (slow cooling)} &  \mrow{2}{20mm}{\cent slow case I (SI)} &  \mcol{4}{c|}{\cent  slow case II$^*$ (SII) }\\
 &  & \mcol{4}{c|}{\cent * In this case $\epsilon_e/\epsilon_B < 1$ is also possible.} \\
\hline
Condition  & $Y(\gc)>1$ & \mcol{4}{c|}{\cent $Y(\gc)<1$  }  \\
\hline
Frequencies & $\gm<\gc<\hgm$ & \mcol{4}{c|}{\cent  ${\rm max}\{\gm,\hgm\}<\gc$ } \\
\hline\hline
\mcol{1}{c}{}&\mcol{1}{c}{}&\mcol{1}{c}{}&\mcol{1}{c}{}&\mcol{1}{c}{}&\mcol{1}{c}{}\\
\end{tabular}
\end{table*}

Fast cooling solutions with  $\gc <\gm$   are divided to several regimes according to the importance of the KN corrections. ``Fast  case I'' (FI) is characterized by $\gm/\hgm<1$ and weak  KN effects.  When $\gm/\hgm\ll 1$, the solution reduces to Thomson as described in \cite{sari01}. 
The region ``fast  case III'' (FIII) with $\gm/\hgm\approx 1$ has moderate KN effects. The solution is  very complicated  in a whole transition region   $1<\gm/\hgm<(\epsilon_e/\epsilon_B)^{1/3}$ due to the appearance of secondary critical LFs 
(see NAS09)\footnote{The physical condition that separates case II from the gap regime located between case II and case III is that $\hgm<\gamma_{\rm self}<\gm<\gamma_0<\widehat{\hgm}$. As the KN effect becomes weaker ($\hgm$ increases while $\gm$ decreases) in the case II, the second critical electron LF $\widehat{\hgm}$ decreases, which eventually becomes smaller than $\gamma_0$ (where $Y(\gamma_0)=1$) and enters the gap regime. In the gap regime, since the SSC cooling is still efficient at $\widehat{\hgm}$ (i.e., $Y(\widehat{\hgm})>1$ since $\widehat{\hgm}<\gamma_0$), this slightly modifies the synchrotron and SSC spectra. Likewise, critical LFs of higher order should transit downward across $\gamma_0$ before asymptotically  reaching the case III ($\gamma_{\rm self}=\gm=\hgm=\widehat{\hgm}=\cdots$). For this reason, NAS09 approximate the gap regime ($\hgm<\gamma_{\rm self}<\gm<\widehat{\hgm}<\cdots<\gamma_0$) by the case III.}.
Following NAS09  we approximate the spectra in this regime assuming that $\gm/\hgm= 1$. Finally, 
 in ``fast  case II'' (FII) in which  $(\epsilon_e/\epsilon_B)^{1/3}<\gm/\hgm$ and the KN effects are strong.

The strong KN regime FII is further divided to  three sub-cases, depending on the relative  strength of the KN effects that depend location of $\g0$ with respect to $\gamma_{\rm self}$ and $\gm$.
FIIc borders FIII. In this regime  $(\epsilon_e/\epsilon_B)^{1/3}<\gm/\hgm<\epsilon_e/\epsilon_B$ corresponding to {$\gamma_{\rm self}<\gm<\g0$}. KN effects are stronger in   FIIb, in which  $\epsilon_e/\epsilon_B<\gm/\hgm<(\epsilon_e/\epsilon_B)^3$ and  correspondingly {$\gamma_{\rm self}<\g0<\gm$}. Finally KN effects are strongest in  FIIa  $(\epsilon_e/\epsilon_B)^3<\gm/\hgm$ with {$\g0<\gamma_{\rm self}<\gm$}. 

There are two spectral slow cooling regimes  depending on whether $Y(\gc)\gtrsim1$ (a weak KN regime which we call ``slow case I'' (SI), corresponding to the ``slow cooling''  in NAS09) or
$Y(\gc)\lesssim1$ (a strong KN regime which we call ``slow case II'' (SII), corresponding to ``dominant synchrotron cooling'' in NAS09). 
{The conditions that define each spectral regime are summarized in Table \ref{tab:spectral regime}.}

\subsection{The Klein-Nishina Limit}
\label{ss:f_KN}

The KN cross-section $\sigma_{_{\rm KN}}$ is often, and in particular in NAS09,  approximated by a Heviside step function with a break at $m_e c^2 = \gamma h \nu'$, where  $h\nu^\prime$ is the seed photon energy measured in the source rest frame (i.e., $\nu^{\prime} = \nu/\Gamma$, where $\nu$ is the seed photon frequency in the observer frame). We modify this approximation introducing a dimensionless factor $f_{_{\rm KN}}< 1$ such that
\beqn
\label{eq:} 
f_{_{\rm KN}}\,m_ec^2=\gamma h\nu^\prime.
\eeqn
This correction factor, $f_{_{\rm KN}} $,  takes into account  the onset of KN corrections at  energies less than $m_e c^2$. 

The common treatment, $f_{_{\rm KN}}=1$, overestimates both the scattered photon energy and the SSC peak flux (or the $Y$ value at the $\nu F_{\nu}$ peak).  In particular, the mean scattered photon energy, weighted by the differential cross section over scattered photon energy and angle, begins to deviate from the approximate value, $\gamma^2 h \nu'$ at  lower photon energies and the exact SSC photon energy is about $1.5$ times smaller than the approximate value at the photon energy $0.2\, m_ec^2$ and the deviation increases at higher energies. 
Additionally, the cross section begins to decrease when $\gamma h\nu^{\prime}\approx 0.1\, m_e c^2$ and this is important 
when computing the average SSC flux by weighting the cross section over scattered photon energy.
The correction factor should depend, of course, on the exact situation and in particular on the electrons' energy power-law index, $p$ and on the typical electron LF, $\gamma_m$.
To take all these into account we
keep the step-function approximation but we use $f_{_{\rm KN}}=0.2$. 
This choice  of $f_{_{\rm KN}}=0.2$ is also supported by the fact that the analytic SSC spectra with $f_{_{\rm KN}}\sim0.2$ tend to agree with the numerically-reproduced SSC spectra that uses the exact cross-section \cite[][ hereafter DP21]{dp21}. 
This modification will affect the critical LFs, such as $\widehat{\gamma}$, $\gamma_{\rm self}$, and $\gamma_0$ in Eqs. \eqref{eq:hat_gamma}--\eqref{eq:gamma_0}. Therefore, the parameters inferred from the spectral modeling depend on the choice of $f_{_{\rm KN}}$ (see  \S \ref{ss:f_KN_1_solution} below). 

\subsection{Spectrum Normalization}
\label{s:normalization}

An additional ingredient required for a comparison with observations is the normalization of the spectrum, which is not explicitly given by NAS09. To normalize the spectrum in the presence of SSC cooling, we need to estimate the isotropic equivalent number of emitting electrons, $\Niso$, by additionally introducing external parameter, the shock radius $R$.

In fast cooling ($\gc<\gm$), the synchrotron spectrum is normalized by the peak flux density at $\nu_{\rm c}\equiv\nu_{\rm syn}(\gc)$:
\beqn
\label{eq:F_nu_c}
F_{\nu_{\rm c}}=\frac{\Niso\, \Gamma\, \gc \,m_e c^2}{ \nu_{\rm c} [1+Y(\gc)]\,t_{\rm obs}} \left(\frac{1+z}{4\pi d_L^2}\right)\  ,
\eeqn
where $Y(\gc)$ takes into account the SSC cooling of electrons. 
In slow cooling ($\gm<\gc$), the synchrotron spectrum is normalized by the peak flux density at $\nu_{\rm m}\equiv\nu_{\rm syn}(\gm)$, which is not influenced by SSC cooling, and thus it is given by \citep{sari98}:
\beqn
\label{eq:F_nu_m}
F_{\nu_{\rm m}}=\frac{\Niso\, P_{\rm syn}(\gm)}{ \nu_{\rm m}} \left(\frac{1+z}{4\pi d_L^2}\right)\  ,
\eeqn
where $P_{\rm syn}(\gamma)=\sigma_{\rm T} c B^2\gamma^2\Gamma^2/(6\pi)$ is the synchrotron power for a single electron with LF $\gamma$.

To normalize the synchrotron spectrum we obtain $\Niso $ by comparing  the magnetic field energy, $(\Gamma B)^2 V/(8\pi) $,  to the energy of the emitting particles, $\Niso\,\Gamma \,f_p\,\gamma_{\rm m}\, m_e c^2$, where $f_p\equiv (p-1)/(p-2)$.  
To estimate the  volume of the emitting region, $V$, we recall that 
the radius of the shock, $R$, and  $\Gamma$ are related to  $t_{\rm obs}$ as
$
R={4\Gamma^2\,c\,t_{\rm obs}}/{(1+z)}\ .
$
The emitting region width
 is $\Delta R =R/(12\Gamma^2)$, thus $V= \pi R^3/(3\Gamma^2)$.
Comparing the two energies by using the ratio  
$\epsilon_e/\epsilon_B$,  we obtain:
\begin{eqnarray}
\label{eq:N_e}
 \Niso= \frac{8c\,t_{\rm obs}^3}{3f_p\, m_e (1+z)^3}\left(\frac{\epsilon_e}{\epsilon_B}\right) \frac{B^2 \Gamma^5}{\gamma_{\rm m}} \ .
\end{eqnarray}
It is important to note that this derivation doesn't assume that the number of emitting electrons equals the number of electron that the blast wave has swept:  $ 4 \pi R^3 n_0/3  $, where $n_0$ is the number density of the ambient matter.
Remarkably, only this part of the analysis depends on GRB afterglow modeling. One can apply our methods to other astronomical system by replacing this argument with another one that is appropriate to the relevant system. 

\subsection{The General Method}
\label{ss:method}
Given a GRB afterglows SED consisting of low-energy (synchrotron at X-ray) and very-high-energy (IC at TeV) components we describe below a method to infer the instantaneous parameters. As the analytic spectra is characterized by several regimes (see \S \ref{ss:model}), one must first identify the relevant spectral regimes. Then we proceed to infer the condition at  the emitting region for each of the relevant spectral regimes.

We  characterize the observed SED using  the peak frequencies and the peak fluxes: $\nus$, the peak synchrotron frequency, $\nuic$, the peak IC frequency,  $F_{\nu_{\rm s}}$ the peak synchrotron flux  and $Y_{\rm obs}\equiv {\nu_{_{\rm IC}}F_{\nu_{_{\rm IC}}}}/{\nu_{\rm s}F_{\nu_{\rm s}}} $, the IC to synchrotron peak luminosity ratio. A fifth  parameter of a different character is the spectral slope (usually above the synchrton frequency) as will be discussed below. 
Among these parameters, $Y_{\rm obs}$ is the most important in determining the potential spectral regimes.  $Y_{\rm obs}\gg1$ is only possible in weak KN regimes where the IC component does not suffer from the KN suppression (FI and SI), whereas $Y_{\rm obs}\ll1$ implies strong KN effects (FII and SII). The intermediate case with $Y_{\rm obs}\sim1$ is rather complicated since it could be realized by more than one spectral regime depending on model parameters (as is the case with GRB 190114C).

Following DP21 we use five model parameters to describe the emitting region: the LF of the shocked material $\Gamma$ (which is smaller by a factor of $\sqrt{2}$ than the shock front LF in the ultra-relativistic limit), the typical LF of the injected electrons $\gm$,  the magnetic field within the shocked region $B$, the ratio between energy-equipartition parameters $\epsilon_e/\epsilon_B$, and the slope of the injected electron energy distribution $p$.  
Among those, within the piecewise linear approximation of the analytic model, $p$ might be determined from the spectral slop 
above the synchrotron peak.  
Assuming that the solution is within a given spectral regime we solve for the other four using the four observational parameters measured above. 
Then, the derived model parameters must be inspected if they satisfy the conditions that define each spectral regime.  

Due to the generic problems of the analytic modelling, the exact choice of observed parameters could be critical to obtain the solution in each spectral regime.  Therefore, one must take into account the potential uncertainties in observed parameters, some of which could lead to relatively large errors in the inferred parameters (see \S \ref{ss:uncertainty}). 

For a given set of parameters we determine the critical electron LFs: $\gamma_{\rm self}$, $\gamma_0$, $\hgm$, or $\hgc$ and calculate the spectrum. In view of the approximate nature of an analytic model with sharp breaks at the transition frequencies 
a $\chi^2$ fit ends up in being useless. Instead, we apply a qualitative fit to the data by visual inspection. 

The analytic model don't allow for self-absorption which might be relevant at the TeV range for low enough $\Gamma$. We don't take this into account when looking for out best fit parameters.  However, we have added numerically the effect of self-absorption to verify that it wasn't too significant, as is the case in most of our results. {We estimate the opacity for two-photon annihilation at given IC energy $\tau_{\gamma\gamma}$ following \citet{dp19}\footnote{{In general, the opacity  depends on the assumed geometry of the emitting region, which may no be trivial and hence require a  detailed simulation.}} by taking into account the difference in dynamical coefficients  used in that paper and here.  We attenuate the IC spectrum by an energy-dependent factor of $(1-e^{-\tau_{\gamma\gamma}})/\tau_{\gamma\gamma}$, which is appropriate for  internal absorption.}
An additional related effect, the emission of the secondary pairs cannot be included without a full detailed simulations. 

Finally we note that we expect a sharp cutoff in the synchrotron spectrum above the maximal synchrotron energy  $\approx m_e c^2/\alpha$ \citep{Guilbert1983,deJager1996,Aharonian2000}, where $\alpha$ is the fine structure constant.  However, typically for the values considered here IC emission is already dominant at this frequency and this cutoff is not relevant. 

\section{Analytic Solutions to GRB 190114C}
\label{s:GRB190114C}

We turn now to the analytic solutions to GRB 190114C based on the analytic SSC model described in \S \ref{s:model}. 
We adopt $p=2.5$ when discussing GRB 190114C so that the synchrotron spectral slope above the peak ($F_{\nu}\propto\nu^{-p/2}$) roughly matches the observed X-ray to GeV flux ratio and thus we are left just with the four model parameters{: $\gm$, $\Gamma$, $B$, and $\epsilon_e/\epsilon_B$}. We take the redshift and the luminosity distance for the source to be $z=0.42$ and $d_L=2.4$ Gpc, respectively.

Following the general methodology described in \S \ref{ss:method}, we identify the characteristic spectral regimes that are relevant for the observation of GRB 190114C in \S \ref{ss:spectral regime}, and derive analytic solutions for fast and slow cooling regimes in \S \ref{ss:early SED}--\ref{ss:f_KN_1_solution}.

\subsection{Identification of Spectral Regimes}
\label{ss:spectral regime}

The GRB 190114C afterglow spectra is characterized by a broad synchrotron peak ranging over two orders of magnitude in photon energy and $Y_{\rm obs}\sim1$.
A significant KN effect is necessary to explain the former condition but it cannot be too strong in order to satisfy the latter one, which precludes  spectral cases FIIa and FIIb.
Moreover, if $\epsilon_e<\epsilon_B$  the IC energy density is smaller than the magnetic field energy density and thus it  results in $Y_{\rm obs}<1$ (\citealt{sari01}; NAS09). Therefore, we use  $\epsilon_e/\epsilon_B\gtrsim1$ in the following analysis of GRB 190114C\footnote{Note that the assumption $\gc<\hgm$ used in  NAS09 holds here. The condition $\hgm<\gc$  results in strong KN suppression where SSC cooling is highly suppressed ($Y_{\rm obs}\ll 1$), which is irrelevant for GRB 190114C.}.

In the fast-cooling regime, the shape around the peak of the synchrotron spectrum becomes relatively flat in the presence of the KN effect since it makes the SSC cooling of low energy electrons more efficient, reducing their synchrotron flux relative to higher energy electrons 
when $\gm/\hgm>1$ 
(\citealt{Derishev2001,Derishev2003,ando08}; NAS09), whereas in the weak KN regime at $\gm/\hgm<1$, the KN effect is insignificant and the IC to synchrotron luminosity ratio is simply determine by the ratio $\epsilon_e/\epsilon_B$
\citep{snp96,sari01}.  Therefore, the strong KN regime FII is favored for GRB 190114C which has a flat synchrotron peak and $Y_{\rm obs}\sim1$.  Among the three sub-classes within FII, cases FIIa and FIIb have the strongest KN effect, which suppresses the $Y_{\rm obs}$ down to much below unity (see Table \ref{tab:spectral regime}). 
Since  $Y_{\rm obs}\sim1$, we focus on FIIc.

Turning now to the two slow cooling regimes we find that within the error range of observational parameters  there is no slow cooling solution with $Y(\gc)<1$. This excludes the strong KN regime SII in which this condition must hold, leaving only the SI regime.

\subsection{The Early-Time Spectrum}
\label{ss:early SED}
We now apply the analytic SSC model described in \S \ref{s:model} with $t_{\rm obs}=90$ s to the early time spectrum taken between $68$ s and $110$ s after the GRB onset. 
We find the representative solutions for afterglow parameters both for fast and slow cooling regimes, and their possible connection in the parameter space will be further discussed in \S \ref{ss:solution track}. For completeness, we demonstrate the characteristic solution that is obtained with the SSC model without a modification on the KN effect later in \S \ref{ss:f_KN_1_solution}.

\subsubsection{Fast-Cooling Solution}
\label{sss:fast}

In fast case IIc, the synchrotron spectrum peaks at  $\nu_0\equiv\nu_{\rm syn}(\gamma_0)$, where $\gamma_0=[(\epsilon_e/\epsilon_B)\gm\hgm]^{1/2}$, which can be related to the observed frequency of the synchrotron peak $\nu_{\rm s}$ as
\beqn
\label{eq:CaseIIc-eq1}
\nu_{\rm s}=\nu_{0}=C_1\,C_2\,\left(\frac{\epsilon_e}{\epsilon_B}\right)\,\frac{\Gamma}{\gm}\  ,
\eeqn
Here, $C_1\equiv q_e/(2\pi m_e c)/(1+z)$ and $C_2\equiv f_{_{\rm KN}} B_{\rm cr}$ are constants defined by $\nu_{\rm syn}(\gamma) =C_1 B\gamma^2\Gamma$ and $\widehat{\gamma}(\gamma)= C_2/(B\gamma^2)$, respectively. Similarly, the frequency of the IC peak can be constrained from observation of $\nuic$:
\beqn
\label{eq:CaseIIc-eq2}
 \nuic=2\gamma_{\rm m}\hgm\,\nu_{\rm m}=2C_1\,C_2\,\Gamma\gm\  ,
\eeqn
Here the peak scattered photon energy is suppressed due to the KN effect ($\hgm<\gm$). The condition for the apparent IC to synchrotron flux ratio reads
\beqn
\label{eq:CaseIIc-eq3}
 Y_{\rm obs}&=&Y(\gm)\,\frac{\nu_{\rm m} F_{\nu_{\rm m}}}{\nu_{0} F_{\nu_{0} }}\nonumber\\
  &
 =&C_2^{\frac{p-2}{2}}(p-2)^{\frac{1}{2}}\left(\frac{\epsilon_e}{\epsilon_B}\right)^{\frac{p-2}{2}}B^{-\frac{p-2}{2}}\gm^{-\frac{3(p-2)}{2}}\  ,
\eeqn
where $Y(\gamma_{\rm m})\approx[(\epsilon_e/\epsilon_B)\,(p-2)]^{1/2}\,(\gm/\hgm)^{-1/2}$ (see Eq. (42) of NAS09), and the factor $F_{\nu_{0}}/F_{\nu_{\rm m}}=(\nu_0/\nu_{\rm m})^{(p-1)/2}$ reflects the fact that the synchrotron luminosity is dominated by electrons with LF $\g0$ ($>\gm$) due to the KN effect.
Additionally, we take into account the normalization of the peak synchrotron flux density:
\beqn
\label{eq:CaseIIc-eq4}
F_{\nu_{\rm s}}&=&F_{\nu_{0}}
=F_{\nu_{\rm c}}\left(\frac{\widehat{\nu_0}}{\nu_{\rm c}}\right)^{-\frac{1}{2}}\left(\frac{\widehat{\nu_{\rm m}}}{\widehat{\nu_{0}}}\right)^{-\frac{p-1}{4}}\left(\frac{\nu_0}{\nu_{\rm m}}\right)^{-\frac{p-1}{2}}\nonumber\\
&=&C_1^{-1}\,C_2^{-\frac{p}{2}}\,C_3\, \left(\frac{\epsilon_e}{\epsilon_B}\right)^{-\frac{p-2}{2}}\Gamma^{5}\,B^{\frac{p+2}{2}}\gm^{\frac{3p-4}{2}}\  ,
\eeqn
where $C_3=2t_{\rm obs}^2c^3/(3\pi f_p d_L^2)/(1+z)^2$ is a numerical constant. Here, the absolute value of $F_{\nu_0}$ is extrapolated from the flux normalization at $\nu_{\rm c}$, which is determined by Eq. \eqref{eq:F_nu_c}: 
\beqn
\label{eq:CaseIIc-eq5-1}
F_{\nu_{\rm c}}=\frac{\Niso\, \Gamma\, \gc \,m_e c^2}{ \nu_{\rm c} Y(\gc)\,t_{\rm obs}} \left(\frac{1+z}{4\pi d_L^2}\right)\  ,
\eeqn
where $\Niso$ is the isotropic equivalent number of electrons given by Eq. \eqref{eq:N_e}, 
and $Y(\gc)=(\epsilon_e/\epsilon_B)^{(4-p)/2}(\gm/\hgm)^{-(2-p)/2}$. We implicitly use the approximation $1+Y(\gc)\sim Y(\gc)$, assuming $Y(\gc)\gg1$.

Finally, the set of four equations \eqref{eq:CaseIIc-eq1},  \eqref{eq:CaseIIc-eq2}, \eqref{eq:CaseIIc-eq3} and \eqref{eq:CaseIIc-eq4} with four model parameters $B$, $\gm$, $\Gamma$, and $\epsilon_e/\epsilon_B$ can be uniquely solved for a given set of observational parameters $\nu_{\rm s}$, $F_{\nu_{\rm s}}$, $Y_{\rm obs}$, and $\nu_{_{\rm IC}}$. However, one must, of course, check if the derived model parameters satisfy the condition for the FIIc spectral regime: $(\epsilon_e/\epsilon_B)^{1/3}<\gm/\hgm<\epsilon_e/\epsilon_B$ and $\gc<\hgm$.
Reducing the  spectrum to four characteristic values is not trivial and there is a significant uncertainty in this process. In the following we have chosen $h\nu_{\rm s}=13^{+12}_{-6}$ keV, $h\nuic=0.45^{+0.15}_{-0.15}$ TeV, the apparent IC to synchrotron luminosity ratio $Y_{\rm obs}=1.0_{-0.15}^{+0.15}$ and the synchrotron peak energy flux $\nu_{\rm s} F_{\nu_{\rm s}}=6.5\times10^{-8}\ {\rm erg\ cm^{-2} \ s^{-1}}$ (corresponding to a peak flux density $F_{\nu_{\rm s}}=2.1$ mJy):
\beqn
\label{eq:CaseIIc-eps}
\frac{\epsilon_e}{\epsilon_B}&=&(p-2)^{\frac{p+2}{6(p-2)}}\,C_1^{-2}\,C_2^{-\frac{4}{3}}\,C_3^{\frac{1}{3}}\,\nu_{\rm s}^{\frac{5}{3}}\,F_{\nu_{\rm s}}^{-\frac{1}{3}}\,Y_{\rm obs}^{-\frac{p+2}{3(p-2)}}\nonumber\\&=&64\ \left(\frac{f_{_{\rm KN}}}{0.2}\right)^{-\frac{4}{3}}\left(\frac{h\nu_{\rm s}}{13{\, \rm keV}}\right)^{\frac{5}{3}}\left(\frac{Y_{\rm obs}}{1.0}\right)^{-3} \left(\frac{F_{\nu_{\rm s}}}{2.1{\, \rm mJy}}\right)^{-\frac{1}{3}},
\eeqn
\vspace{-0.3cm}
\beqn
\gm &=&2^{-\frac{1}{2}}\,(p-2)^{\frac{p+2}{12(p-2)}}\,C_1^{-1}\,C_2^{-\frac{2}{3}}\,C_3^{\frac{1}{6}}\,\nu_{\rm s}^{\frac{1}{3}}\,\nu_{_{\rm IC}}^{\frac{1}{2}}\,F_{\nu_{\rm s}}^{-\frac{1}{6}}\,Y_{\rm obs}^{-\frac{p+2}{6(p-2)}}\nonumber\\&=& 3.3\times10^4\left(\frac{f_{_{\rm KN}}}{0.2}\right)^{-\frac{2}{3}}\left(\frac{h\nu_{\rm s}}{13{\, \rm keV}}\right)^{\frac{1}{3}}\left(\frac{h\nu_{_{\rm IC}}}{0.45{\, \rm TeV}}\right)^{\frac{1}{2}}\\&&\times\left(\frac{Y_{\rm obs}}{1.0}\right)^{-\frac{3}{2}}\left(\frac{F_{\nu_{\rm s}}}{2.1{\, \rm mJy}}\right)^{-\frac{1}{6}},
\nonumber
\eeqn
\vspace{-0.3cm}
\beqn
B &=&2^{\frac{3}{2}}\,(p-2)^{-\frac{p-10}{12(p-2)}}\,C_1\,C_2^{\frac{5}{3}}\,C_3^{-\frac{1}{6}}\,\nu_{\rm s}^{\frac{2}{3}}\,\nu_{_{\rm IC}}^{-\frac{3}{2}}\,F_{\nu_{\rm s}}^{\frac{1}{6}}\,Y_{\rm obs}^{\frac{p-10}{6(p-2)}}\nonumber\\&=& 3.9\ {\rm G}\ \left(\frac{f_{_{\rm KN}}}{0.2}\right)^{\frac{5}{3}}\left(\frac{h\nu_{\rm s}}{13{\, \rm keV}}\right)^{\frac{2}{3}}\left(\frac{h\nu_{_{\rm IC}}}{0.45{\, \rm TeV}}\right)^{-\frac{3}{2}}\\&&\times\left(\frac{Y_{\rm obs}}{1.0}\right)^{-\frac{5}{2}}\left(\frac{F_{\nu_{\rm s}}}{2.1{\, \rm mJy}}\right)^{\frac{1}{6}},\nonumber
\eeqn
\vspace{-0.3cm}
\beqn
\label{eq:CaseIIc-Gamma}
\Gamma &=&2^{-\frac{1}{2}}\,(p-2)^{-\frac{p+2}{12(p-2)}}\,C_2^{-\frac{1}{3}}\,C_3^{-\frac{1}{6}}\,\nu_{\rm s}^{-\frac{1}{3}}\,\nu_{_{\rm IC}}^{\frac{1}{2}}\,F_{\nu_{\rm s}}^{\frac{1}{6}}\,Y_{\rm obs}^{\frac{p+2}{6(p-2)}}\nonumber\\&=& 94\ \left(\frac{f_{_{\rm KN}}}{0.2}\right)^{-\frac{1}{3}}\left(\frac{h\nu_{\rm s}}{13{\, \rm keV}}\right)^{-\frac{1}{3}}\left(\frac{h\nu_{_{\rm IC}}}{0.45{\, \rm TeV}}\right)^{\frac{1}{2}}\\&&\times\left(\frac{Y_{\rm obs}}{1.0}\right)^{\frac{3}{2}}\left(\frac{F_{\nu_{\rm s}}}{2.1{\, \rm mJy}}\right)^{\frac{1}{6}},\nonumber
\eeqn
where $p=2.5$ 
have been used in the 
second equality of each equation. 
Figure \ref{fig:YP21fast} shows the spectrum for our analytic fast-cooling solution ({F2C} hereafter). 
One can see that the overall spectrum seems to be in good agreement with the observation.

For completeness we present below the spectral slopes  in FIIc which is most likely the relevant regime (see  NAS09 for details and for the spectral slopes in other regimes). In this regime the synchrotron part of the spectrum (see Fig. \ref{fig:YP21fast}) is  modified at $\widehat{\nu_0}<\nu<\nu_0$ from the original spectrum without KN effects  \cite[i.e., ][]{sari01}. Note that $\widehat{\nu_0}$ is too low to be shown in Fig. \ref{fig:YP21fast}.
It is characterized as a broken power law with :
\beqn
\nu F_\nu \propto \begin{cases}
\nu^{\frac{4}{3}} &  {\rm for }~ \nu<\nu_{\rm c} \ , \\
\nu^{\frac{1}{2}} & {\rm for }~ \nu_{\rm c}<\nu <\widehat{\nu_{\rm 0}} \ , \\
\nu^{-\frac{p-5}{4}} & {\rm for }~\widehat{\nu_0}<\nu <\widehat{\nu_{\rm m}}~{\rm (modified)} \ , \\
\nu & {\rm for }~\widehat{\nu_{\rm m}}<\nu<\nu_{\rm m}~{\rm (modified)} \ , \\
\nu^{-\frac{p-3}{2}} & {\rm for }~\nu_{\rm m}<\nu <\nu_0~{\rm (modified)} \ , \\
\nu^{-\frac{p-2}{2}} & {\rm for }~\nu_0<\nu \ .
 \end{cases}
\eeqn
The IC part of the spectrum is described by:
\beqn
\nu F_\nu \propto 
\begin{cases}
\nu^{\frac{3}{4}} & {\rm for }~\nu<2\nu_{\rm m}\hgm\gm~{\rm (modified)}\ ,\\
\nu^{-\frac{2p-3}{2}} & {\rm for }~2\nu_{\rm m}\hgm\gm<\nu~{\rm (modified)} \ .
\end{cases}
\eeqn

\begin{figure}
\centering
\includegraphics[width=.5\textwidth]{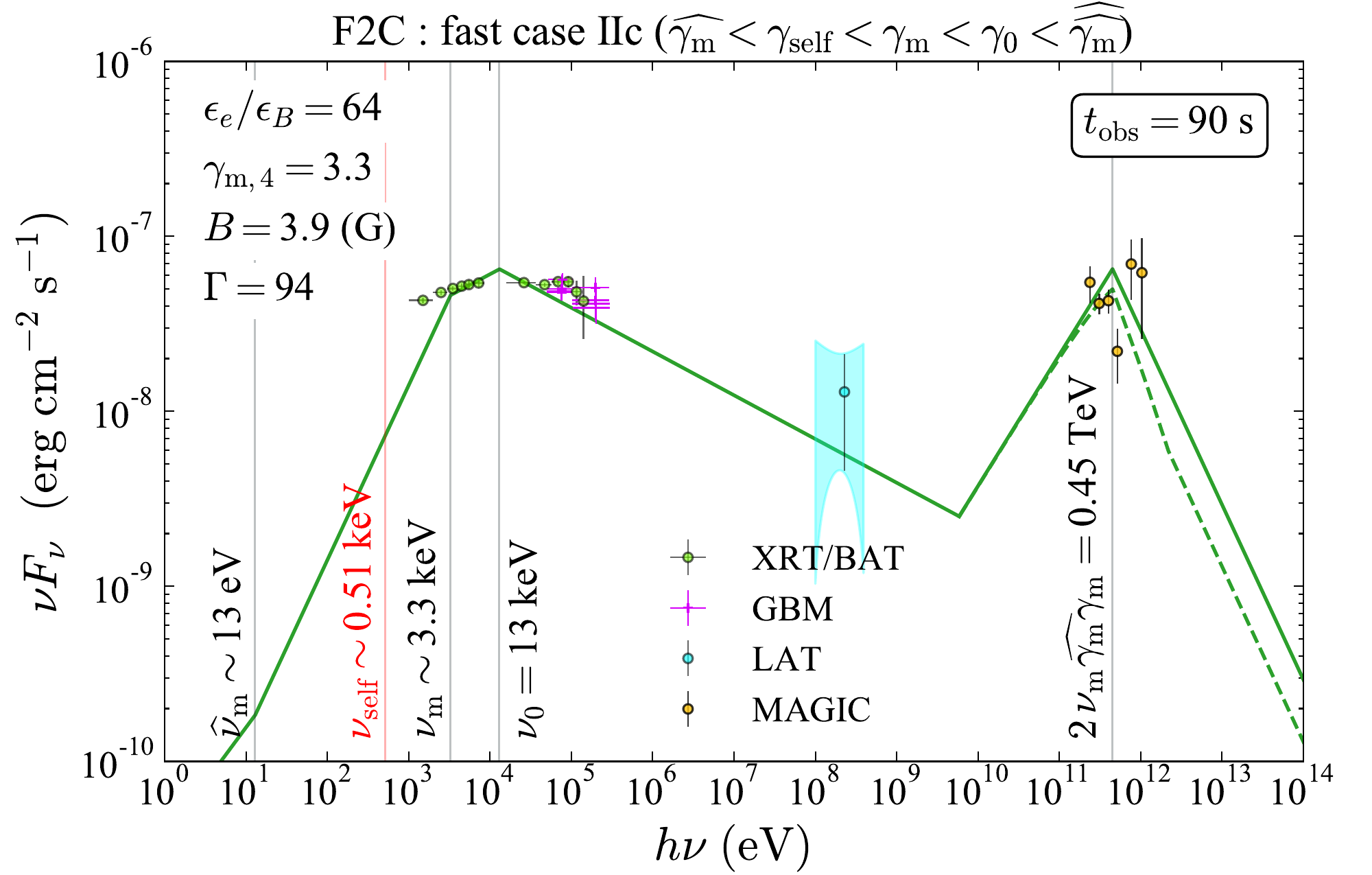}
  \caption{A comparison of the analytic fast-cooling  solution (F2C) with the observed multi-wavelength (X-ray, GeV and TeV) spectrum of GRB 190114C at $t_{\rm obs}=90$ s \citep{magic19,ajello20}. The theoretical spectrum is for the analytic fast-cooling solution obtained by requiring $h\nu_{\rm s}=13$ keV, $h\nu_{_{\rm IC}}=0.45$ TeV, $Y_{\rm obs}=1.0$, $\nus F_{\nus}=6.5\times10^{-8}\ {\rm erg\ cm^{-2} \ s^{-1}}$ with $f_{_{\rm KN}}=0.2$ and $p=2.5$. {The inferred model parameters are shown on the top left and the key frequencies are  indicated with thin lines.} Other relevant parameters calculated internally are: $\gamma_{\rm c}^{\rm syn}=8.8\times10^3$, $\gc=1.9\times10^2$, $\hgm=2.1\times10^3$, $\gamma_{\rm self}=1.3\times10^4$, $\gamma_0=6.6\times10^4$, $\widehat{\hgm}=5.3\times10^5$, and $\widehat{\gamma_0}=5.2\times10^2$. {The spectrum with self absorption is  shown in dashed line.}
  }
\label{fig:YP21fast}
\end{figure}

\begin{figure*}
\centering
\includegraphics[width=0.93\textwidth]{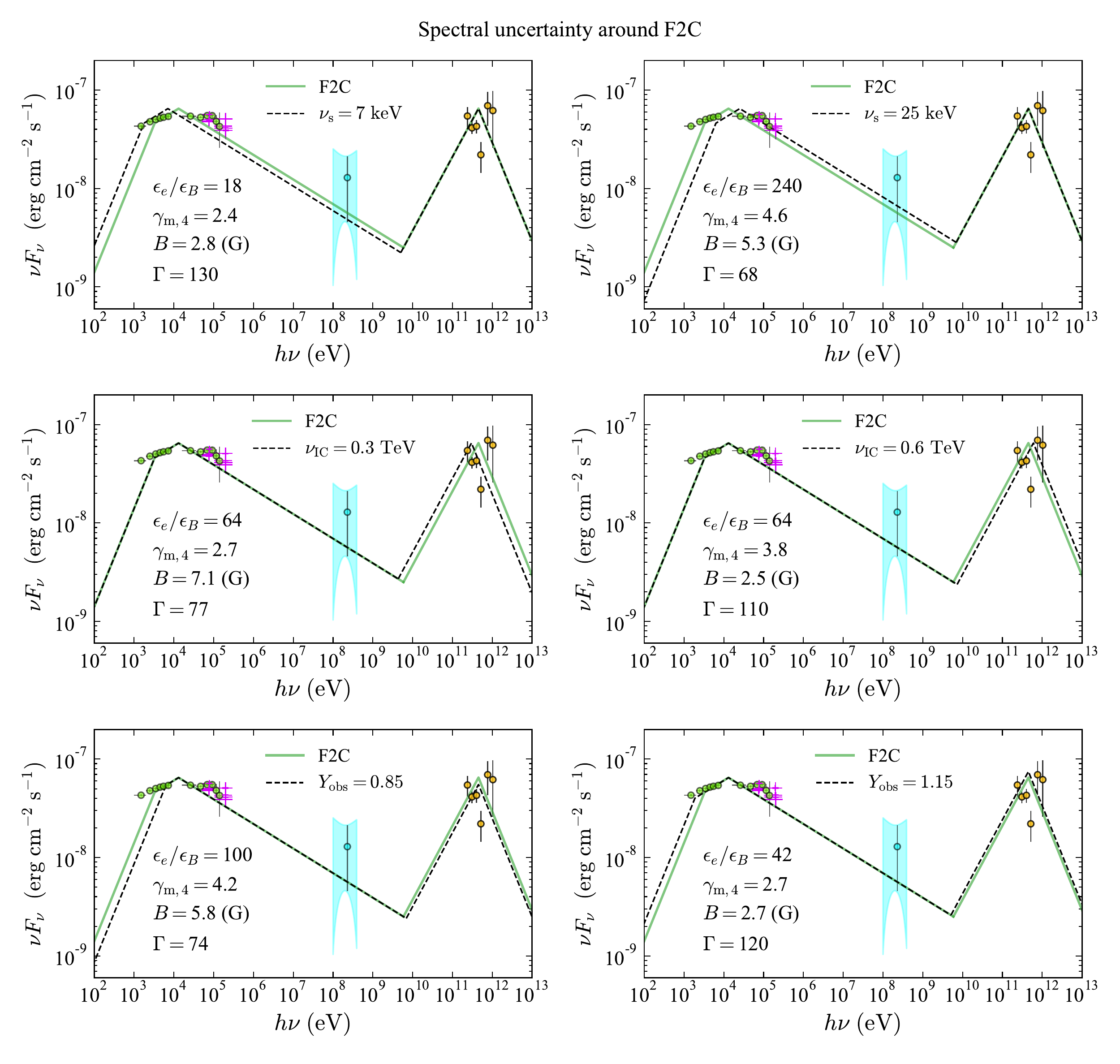}
 \caption{Spectra (black dashed lines) and the inferred parameters (indicated in each panel) when varying observational parameters describing the spectrum, $\nus$ (top panels), $\nuic$ (middle panels), and $Y_{\rm obs}$ (bottom panels), from the canonical values used to determine F2C 
 (green solid lines which are the same as Figure \ref{fig:YP21fast}), within the error range of the observed quantities.}
\label{fig:spectral_error}
\end{figure*}

\subsubsection{Uncertainty in Parameter Inference}
\label{ss:uncertainty}
The inferred parameters depend strongly on some the observed quantities, or more specifically on the quantities we use to characterize the spectrum. 
The dependence on  the IC parameters and in particular on $Y_{\rm obs}$ is the strongest, while the dependence on the synchrotron parameters $F_{\nu_{\rm s}}$ and $\nu_{\rm s}$ is weaker. 
Within the  FIIc solution, the model parameter $\gm$,  which defines the spectral regime of the system, has an exceptionally strong dependence on $Y_{\rm obs}$, with $\gm/\hgm\propto Y_{\rm obs}^{-7}$.  For example, if one adopts $Y_{\rm obs}\gtrsim 1.3$, (just slightly above our error range $Y_{\rm obs}=1 \pm 0.15$) the  spectral regime shifts from FIIc to FIII and the above  solution is no longer relevant.

Figure \ref{fig:spectral_error} shows  the dependence of the  solution F2C  on  the observational parameter.  Even though the spectral shape is almost the same, the inferred parameters change significantly. We have also  estimated the errors of inferred parameters of F2C by a Monte Carlo method assuming that the observational parameters ($\nus$, $\nuic$, $Y_{\rm obs}$) follow a log-normal distribution with its $2$--$\sigma$ error being set by the error range we choose: $h\nus/$(keV)$\,\in[7,25]$ , $h\nuic/$(TeV)$\,\in[0.3,0.6]$, and $Y_{\rm obs}\in[0.85,1.15]$ (the peak flux is set to be a canonical value since its variation barely affects the result). The uncertainty around the solution is estimated via Eqs. \eqref{eq:CaseIIc-eps}--\eqref{eq:CaseIIc-Gamma} as $\epsilon_e/\epsilon_B=64_{-47}^{+196}$, $\gamma_{\rm m}=3.3_{-1.2}^{+1.8}\times10^4$, $B=3.9_{-1.8}^{+5.0}$ G, and $\Gamma=94_{-36}^{+46}$.
Clearly, the uncertainty around $\epsilon_e/\epsilon_B$ is the largest up to a factor of $4$ due to the relatively strong dependence on $\nus$ and $Y_{\rm obs}$ as seen in Eq. \eqref{eq:CaseIIc-eps}.

\subsubsection{Slow-Cooling Solution}
\label{sss:slow}

In the SI spectral regime, the peak of the synchrotron spectrum can have a complicated structure depending on the order of characteristic frequencies around $\gamma_{\rm self}$. 
Here, we specifically consider the case of $\gamma_0>\hgm$.
Similarly to the FIIc case, we can solve (although numerically) four equations on $\nu_{\rm s}$, $F_{\nu_{\rm s}}$, $Y_{\rm obs}$, and $\nu_{_{\rm IC}}$ for $\epsilon_e/\epsilon_B$, $\gm$, $B$, $\Gamma$.

The synchrotron spectrum peaks at  $\nu_0\equiv\nu_{\rm syn}(\g0)$, where $\g0= Y_{\rm c}^{\frac{3}{4}}\hgm^{\frac{3p-1}{8}}\left[{\rm max}(\gc,\hgc)\right]^{-\frac{3(p-3)}{8}}$ with $Y_{\rm c}\equiv Y(\gc)$:
\beqn
\label{eq:SI-eq1}
&&\nu_{\rm s}=\nu_{0}=C_1\,Y_{\rm c}^{\frac{3}{2}}\gm^{-\frac{3p-1}{2}}\,\Gamma\nonumber\\&&\times\begin{cases}
    C_2^2\,\gc^{\frac{3(p-3)}{2}}B^{-1}& (\gc<\hgc),\\
    C_2^{-\frac{3p-1}{4}}\,\gc^{-\frac{3(p-3)}{4}}B^{-\frac{3p-5}{4}}& (\gc>\hgc).
\end{cases}
\eeqn
Using the observed peak of the high energy emission we have: 
\beqn
\label{eq:SI-eq2}
 \nu_{_{\rm IC}}=2\nu_{\rm c}\gc\, {\rm min}(\gc,\hgc)=\begin{cases}
    2C_1\,\gc^4\,B\,\Gamma& (\gc<\hgc),\\
    2C_1\,C_2\,\gc\,\Gamma& (\gc>\hgc).
\end{cases}
\eeqn
Here ${\rm min}(\gc,\hgc)$ reflects the KN suppression. The apparent IC to synchrotron flux ratio is given by: 
\begin{eqnarray}
\label{eq:SI-eq3}
 &&Y_{\rm obs}=Y_{\rm c}\,\frac{\nu_{\rm c}F_{\nu_{\rm c}}}{\nu_{0}F_{\nu_{0}}}=Y_{\rm c}^{\frac{3p-6}{4}}\gm^{-\frac{3p^2-7p+2}{4}}\nonumber\\&&\times\begin{cases}
    C_2^{p-2}\,\gc^{\frac{3p^2-19p+26}{4}}B^{-p+2}& (\gc<\hgc),\\
    C_2^{\frac{3p^2-7p+2}{8}}\,\gc^{-\frac{3p^2-7p+2}{8}}B^{-\frac{3p^2-7p+2}{8}}& (\gc>\hgc).
    \end{cases}
\end{eqnarray}
Additionally, we take into account the normalization of the peak synchrotron flux density:
\beqn
\label{eq:SI-eq4}
&&F_{\nu_{\rm s}}=F_{\nu_{0}}=F_{\nu_{\rm m}}
\nonumber\\&&\times\begin{cases}
    \left(\frac{\nu_{\rm c}}{\nu_{\rm m}}\right)^{-\frac{p-1}{2}}\left(\frac{\widehat{\nu_{\rm c}}}{\nu_{\rm c}}\right)^{-\frac{p}{2}}\left(\frac{\widehat{\nu_{\rm m}}}{\widehat{\nu_{\rm c}}}\right)^{-\frac{3(p-1)}{4}}\left(\frac{\nu_0}{\widehat{\nu_{\rm m}}}\right)^{-\left(\frac{p}{2}-\frac{2}{3}\right)}& (\gc<\hgc),\\
    \left(\frac{\nu_{\rm c}}{\nu_{\rm m}}\right)^{-\frac{p-1}{2}}\left(\frac{\widehat{\nu_{\rm m}}}{\nu_{\rm c}}\right)^{-\frac{3(p-1)}{4}}\left(\frac{\nu_0}{\widehat{\nu_{\rm m}}}\right)^{-\left(\frac{p}{2}-\frac{2}{3}\right)}& (\gc>\hgc),
    \end{cases}\nonumber\\
&&=C_1^{-1}C_4\,Y_{\rm c}^{\frac{-3p+4}{4}}\left(\frac{\epsilon_e}{\epsilon_B}\right)\Gamma^6\nonumber\\
&&\times\begin{cases}
    C_2^{-p}\gc^{\frac{-3p^2+9p+4}{4}}\gm^{\frac{3p^2+3p-8}{4}}B^{3+p}& (\gc<\hgc),\\
    C_2^{\frac{-3p^2+p}{8}}\gc^{\frac{3p^2-5p+4}{8}}\gm^{\frac{3p^2+p-6}{4}}B^{\frac{3p^2-p+24}{8}}& (\gc>\hgc),
    \end{cases}
\eeqn
where $F_{\nu_{0}}$ has been extrapolated from the flux normalization at $\nu_{\rm m}$, which is determined by Eq. \eqref{eq:F_nu_m}: 
\beqn
F_{\nu_{\rm m}}=\frac{\Niso\, P_{\rm syn}(\gm)}{ \nu_{\rm m}} \left(\frac{1+z}{4\pi d_L^2}\right)
=C_1^{-1}C_4\,B^3\gm^{-1}\left(\frac{\epsilon_e}{\epsilon_B}\right)\Gamma^6,
\eeqn
where $C_4=\sigma_{\rm T}c^2 t_{\rm obs}^3/(9\pi^2f_p m_e d_L^2)/(1+z)^2$ is a numerical constant. Lastly, there are internal conditions for $\gc$ and $Y_{\rm c}$:
\beqn
\label{eq:SI-eq5}
 \gamma_{\rm c}\,(1+Y_{\rm c})=\gamma_{\rm c}^{\rm syn}=\frac{C_5}{B^2\Gamma},
 \eeqn
 \vspace{-0.3cm}
 \beqn
 \label{eq:SI-eq6}
Y_{\rm c}(1+Y_{\rm c})&=&\left(\frac{\epsilon_e}{\epsilon_B}\right)\left(\frac{\gc}{\gm}\right)^{2-p} \left(\frac{{\rm min}(\gc,\hgc)}{\gc}\right)^{\frac{3-p}{2}}\\
&=&\left(\frac{\epsilon_e}{\epsilon_B}\right)\gm^{p-2}\begin{cases}
    \gc^{-p+2}& (\gc<\hgc),\\
    C_2^{-\frac{p-3}{2}}\gc^{\frac{p-5}{2}}B^{\frac{p-3}{2}}& (\gc>\hgc),
    \end{cases}\nonumber
\eeqn
where Eq. \eqref{eq:SI-eq5} is identical to Eq. \eqref{eq:gc} and $C_5 = 6\pi m_ec(1+z)/(\sigma_{\rm T}t_{\rm obs})$ is a numerical constant. 

\begin{figure}
\centering
\includegraphics[width=.5\textwidth]{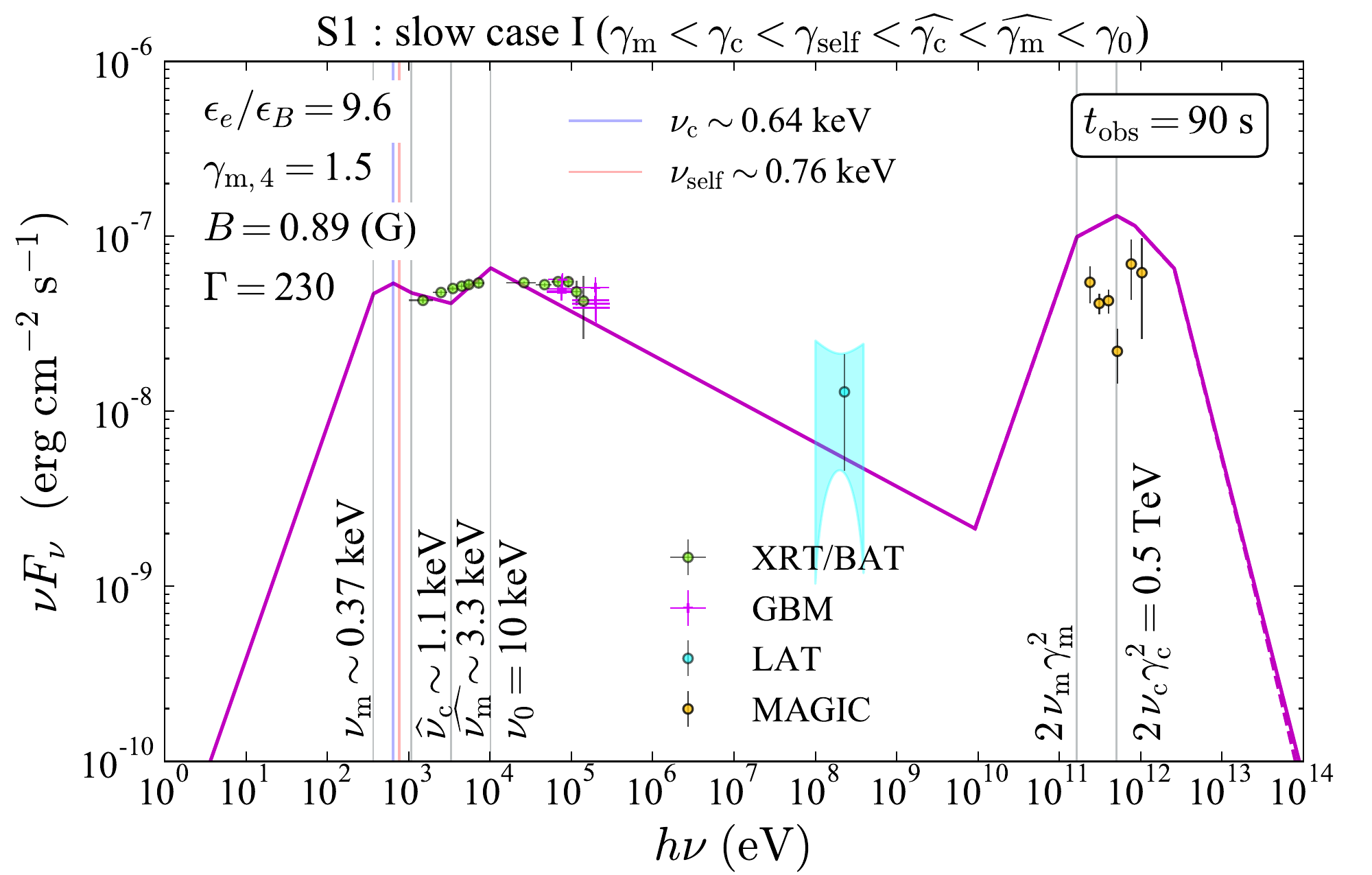}
 \caption{The best fit analytic slow-cooling solution (S1) obtained by requiring  $h\nu_{\rm s}=10$ keV, $h\nu_{_{\rm IC}}=0.50$ TeV, $\nu_{\rm s} F_{\nu_{\rm s}}=6.6\times10^{-8}\ {\rm erg\ cm^{-2} \ s^{-1}}$, $Y_{\rm obs}=2.0$ with $f_{_{\rm KN}}=0.2$ and $p=2.5$.  
 The spectrum with self absorption is  shown in dashed line {although it is almost indiscernible due to the large $\Gamma$}.
 Note that $Y_{\rm obs} =2.0$ here as there is no slow-cooling solution for lower values. 
 }
\label{fig:YP21slow}
\end{figure}

For a given set of observational parameters $\nu_{\rm s}$, $F_{\nus}$, $Y_{\rm obs}$, and $\nu_{_{\rm IC}}$, one can uniquely solve six equations \eqref{eq:SI-eq1}, \eqref{eq:SI-eq2}, \eqref{eq:SI-eq3}, \eqref{eq:SI-eq4}, \eqref{eq:SI-eq5} and \eqref{eq:SI-eq6} for the four external model parameters $\epsilon_e/\epsilon_B$, $\gm$, $B$ and $\Gamma$, and  two additional internal parameters $\gc$ and $Y_{\rm c}$. 
Since those equations are non-linear, we solve them numerically. Similarly to the fast cooling case, the derived model parameters must be checked if they satisfy the conditions that define the SI spectral regime: $\gc>\gm$ and $Y_{\rm c}>1$. 
Somewhat surprisingly, there is no slow cooling solution for the canonical set of parameters $h\nus=13$ keV, $h\nuic=0.45$ TeV, $Y_{\rm obs}=1.0$, and $F_{\nus}=2.1$ mJy that we have chosen to describe the spectrum. 
In particular, for these values of  $\nus$, $\nuic$ and $F_{\nus}$ within the observational uncertainty, there are SI solutions only if $Y_{\rm obs}\gtrsim 1.8$.
Alternatively solutions with $Y_{\rm obs} \lesssim1.8$ are only possible if the value of $\nus$ is below the observed uncertainty range that we defined. Such solutions, however, are inconsistent with the X-ray observations\footnote{An alternative possibility is to look for  SI solutions with lower $\nuic$ and higher $Y_{\rm obs}$ keeping the remaining observational parameters ($\nus$ and $\nu_s F_{\nus}$) so that the ratio between the IC flux at TeV $\nu F_{\nu}(0.5\,{\rm TeV})$ and synchrotron peak flux $\nus F_{\nus}$ becomes close to unity.  
However, there are no slow cooling solutions with $\nuic\lesssim0.1$/ TeV. Within the SI regime with $\nuic\gtrsim0.1$ TeV the minimal possible value of  $\nu F_{\nu}(0.5\,{\rm TeV})/(\nus F_{\nus})$ is  $\approx 1.8$. }.

To explore, in spite of this fact, the slow cooling regime, we have adopted slightly different observational values for SI than the canonical ones adopted for FIIc (see \S \ref{sss:fast}). 
Choosing $h\nu_{\rm s}=10$ keV, $h\nu_{_{\rm IC}}=0.50$ TeV, $\nu_{\rm s} F_{\nu_{\rm s}}=6.6\times10^{-8}\ {\rm erg\ cm^{-2} \ s^{-1}}$, $Y_{\rm obs}=2.0$ with $f_{_{\rm KN}}=0.2$ and $p=2.5$, we obtain a solution with $\epsilon_e/\epsilon_B\sim9.6$, $\gm\sim1.5\times10^4$, $B\sim0.89$ G, $\Gamma\sim2.3\times10^2$, $\gc\sim2.0\times10^4$, and $Y_c\sim2.4$ in the regime of $\gc<\hgc$. 
Figure \ref{fig:YP21slow} depicts the spectrum for this analytic slow-cooling solution ({S1} hereafter). 
{The relatively broad peak in the synchrotron component explains the observations well, but there is a clear excess in the IC component due to the choice of the large value of $Y_{\rm obs}$.  This cannot be mitigated even when accounting for the internal absorption due to the large value of $\Gamma$ ({see the dashed line in Figure \ref{fig:YP21slow}}). 
Slightly different choices of $Y_{\rm obs}$ and $\nuic$, all lead to comparably poor fit quality at the TeV range.  This S1 is just a representative case among such solutions.
Clearly fast cooling solutions are preferred over slow cooing ones.  However, since our approach is only qualitative, we accept the discrepancy of S1 in the IC component to see how it relates to the  fast cooling solutions  in \S \ref{ss:solution track}. }

\begin{figure}
\centering
\includegraphics[width=.5\textwidth]{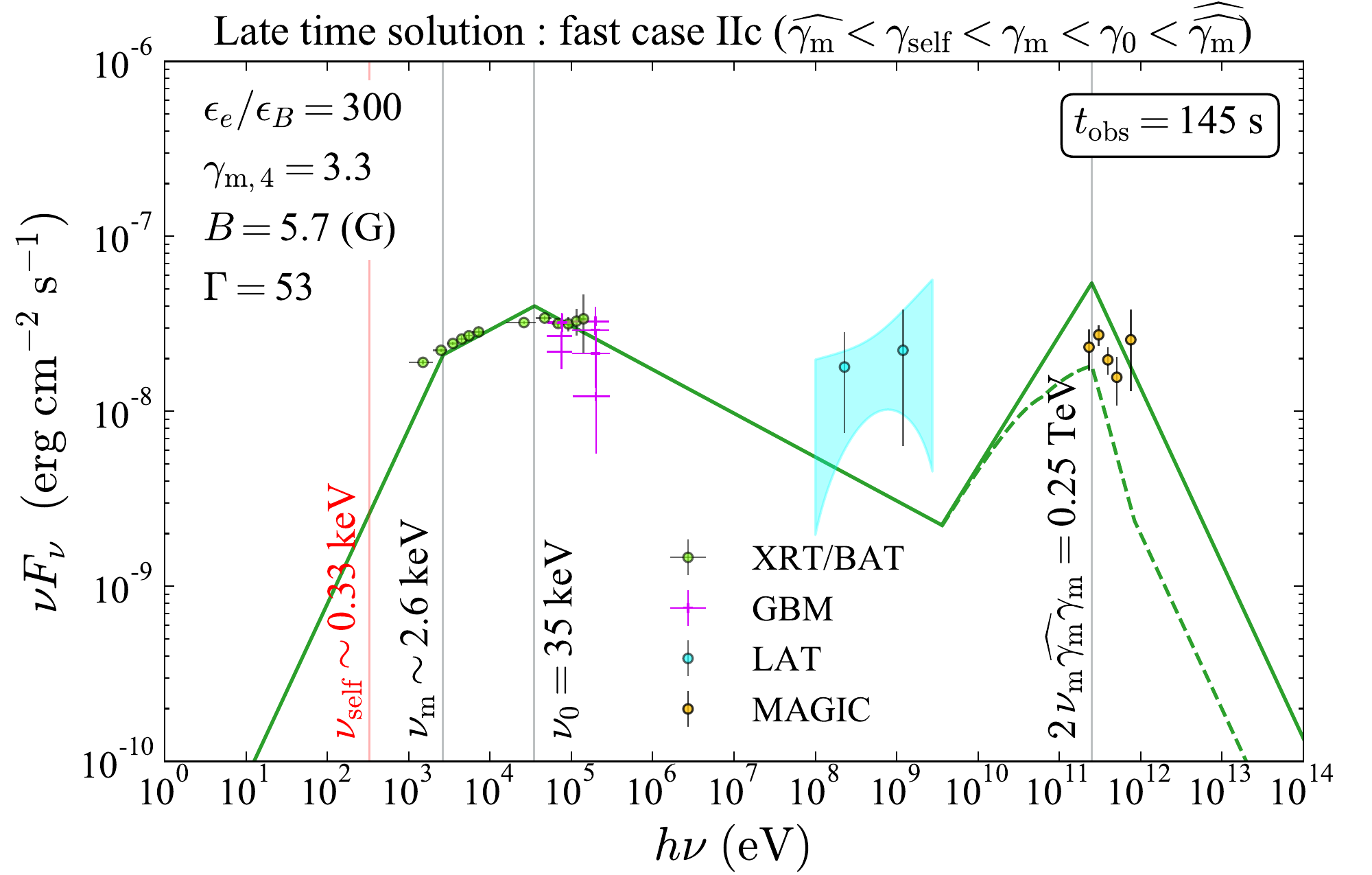}
  \caption{Same as Figure \ref{fig:YP21fast} but for the late time spectrum of GRB 190114C at $t_{\rm obs}=145$ s \citep{magic19}. The spectrum is for the analytic fast-cooling solution obtained by requiring $h\nu_{\rm s}=35$ keV, $h\nu_{_{\rm IC}}=0.25$ TeV, $Y_{\rm obs}=1.35$, $\nu_{\rm s}F_{\nu_{\rm s}}=4.0\times10^{-8}\ {\rm erg\,cm^{-2}\,s^{-1}}$ with $f_{_{\rm KN}}=0.2$ and $p=2.5$. 
 {The spectrum with self absorption is  shown in dashed line.}
  }
\label{fig:YP21fastlate}
\end{figure}

\subsection{The Late-Time Spectrum}
\label{ss:late SED}
We apply the same procedure to the  $t_{\rm obs}=145$ s to approximate the late time spectrum taken between $110$ s and $180$ s after the GRB onset. 
We choose as the observables characterizing the spectrum:  $h\nu_{\rm s}=35_{-15}^{+25}$ keV, $h\nuic=0.25_{-0.1}^{+0.1}$ TeV, $Y_{\rm obs}=1.35_{-0.15}^{+0.15}$ and $\nu_{\rm s} F_{\nu_{\rm s}}=4.0\times10^{-8}\ {\rm erg\ cm^{-2} \ s^{-1}}$ with $p=2.5$.
We only discuss the fast cooling spectrum since unlike the early time analysis we find no slow cooling solution with acceptable fit to the given data set.  
Assuming that the spectrum is in the FIIc spectral regime, 
we obtain a  solution with {$\epsilon_e/\epsilon_B=300_{-200}^{+640}$, $\gm=3.3_{-1.1}^{+1.3}\times10^4$, $B=5.7_{-2.5}^{+7.3}$ G, and $\Gamma=53_{-18}^{+20}$, where the $2$--$\sigma$ uncertainty around the solution is estimated by the same method as used for the early time solution in \S \ref{ss:uncertainty}.

Figure \ref{fig:YP21fastlate} shows a comparison between the model spectrum  and the late time data. Except for the slightly insufficient flux at GeV, the quality of the fit is as good as the early time result. Table \ref{tab:summary} summarizes the fast-cooling solutions for early and late times. 
Given the large uncertainty in the inferred parameters the solutions for early and late times could in principle have the same model parameters. The evolution of the physical parameters is not robustly constrained because the uncertainties are large.
Nevertheless, when looking at the canonical values one can observe the following trends. First, the bulk LF $\Gamma$ has decreased from $94$ to $53$, which may reflect the expected deceleration of the source. Secondly, the ratio $\gm/\hgm\propto B\gm^3 $ increases by about $40$\% from early to late, implying that the KN effect becomes stronger at later times. Furthermore, the two solutions suggest that $\epsilon_e/\epsilon_B$ increases in time from $64$ to $300$ and $B$ increases from $3.9$ G to $5.7$ G. 
Assuming that the ambient matter density is constant (i.e., $\epsilon_B\propto B^2/\Gamma^2$), the temporal increase in $\epsilon_B$ is estimated to be about $6.7$ times.
This, combined with the change in $\epsilon_e/\epsilon_B$, implies that $\epsilon_e$ has increased by a factor of $31$.

\begin{table}
\centering
\caption{Summary of characteristic afterglow parameters inferred from fits of the analytic synchrotron-SSC model to GRB 190114C. The parameters below the second double lines are fixed during the fit procedure. The parameter set for the early time fit ($t_{\rm obs}=90$ s) corresponds to F2C solution. }
\label{tab:summary}
\begin{tabular}{lccc}
\hline\hline
   & $t_{\rm obs}=90$ s&    & $t_{\rm obs}=145$ s      \\ \cline{2-2}\cline{4-4}
$\Gamma$                & $94_{-36}^{+46}$           &      & $53_{-18}^{+20}$                     \\           
$B$                  & $3.9_{-1.8}^{+5.0}$ G          &        & $5.7_{-2.5}^{+7.3}$ G                \\ 
$\gm$                   & $3.3_{-1.2}^{+1.8}\times10^4$  & &$3.3_{-1.1}^{+1.3}\times10^4$  \\ 
$\epsilon_e/\epsilon_B$ & $64_{-47}^{+196}$            &       & $300_{-200}^{+640}$                   \\ \hline\hline
$p$                     & \multicolumn{3}{c}{$2.5$}        \\ 
$f_{_{\rm KN}}$                     & \multicolumn{3}{c}{$0.2$}       \\ 
Spectral regime         & \multicolumn{3}{c}{fast case IIc} 
\\ 
Section        & \S \ref{sss:fast}  &       & \S \ref{ss:late SED} \\ \hline\hline

\end{tabular}
\end{table}

\begin{figure*}
\centering
\includegraphics[width=0.96\textwidth]{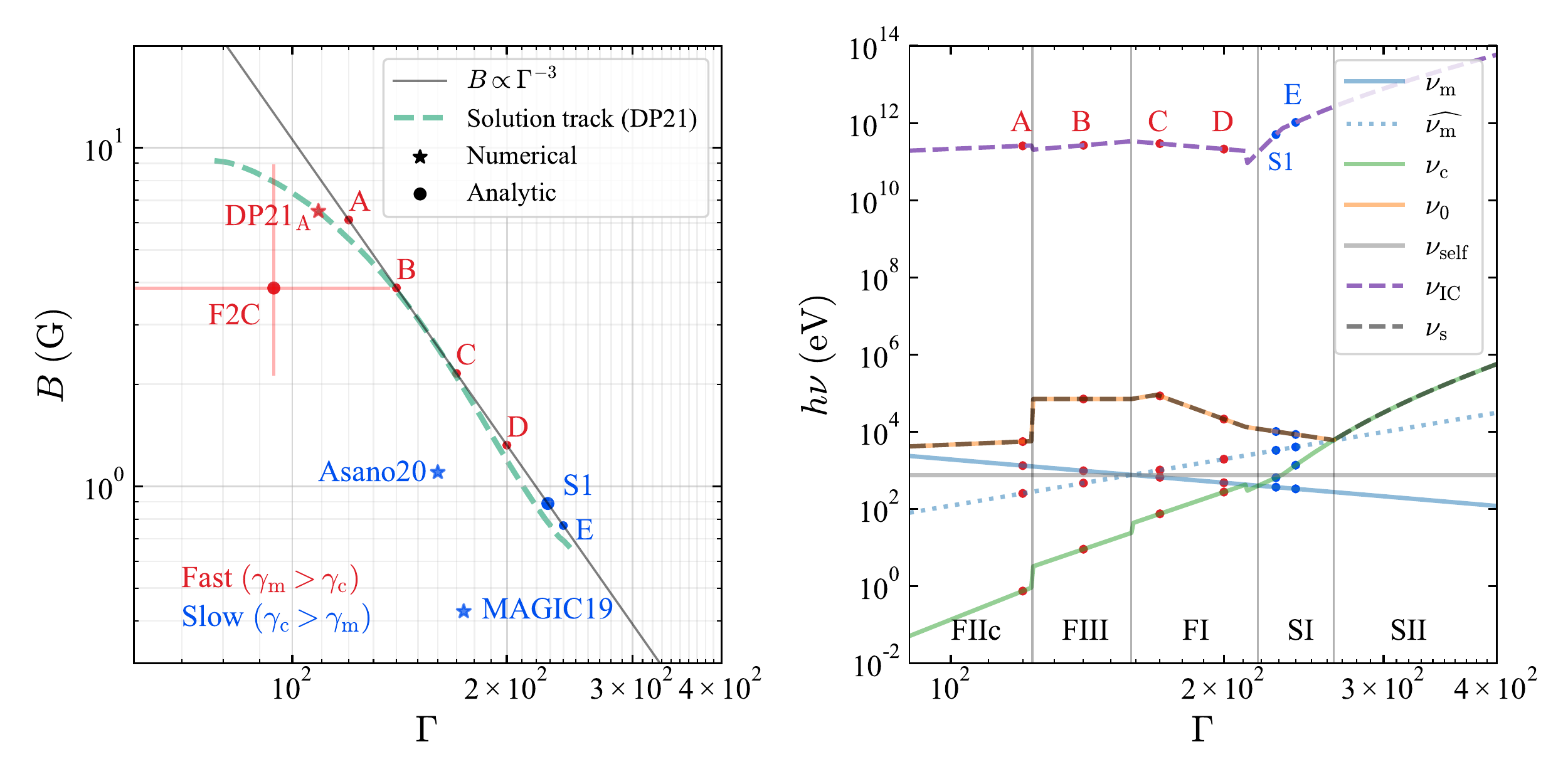}
 \caption{A solution track in $\Gamma$--$B$ plane.
 {\it
    Left}: Analytic  and  numerical solutions  on the $\Gamma$--$B$ plane.  {\it Right}:  the variation of characteristic break frequencies along the $B\propto\Gamma^{-3}$ track (with $\epsilon_e/\epsilon_B=9.6$ and $\gm=1.5\times10^4$ fixed) that passes the analytic slow cooling solution. The vertical lines represent the boundaries between different spectral regimes (see \S \ref{ss:model} for details).  Each dot corresponds to the representative set of parameters shown in Figure \ref{fig:spectra_mosaic}.  }
\label{fig:Gamma_B_map_summary}
\end{figure*}

\begin{figure*}
\centering
\includegraphics[width=1\textwidth]{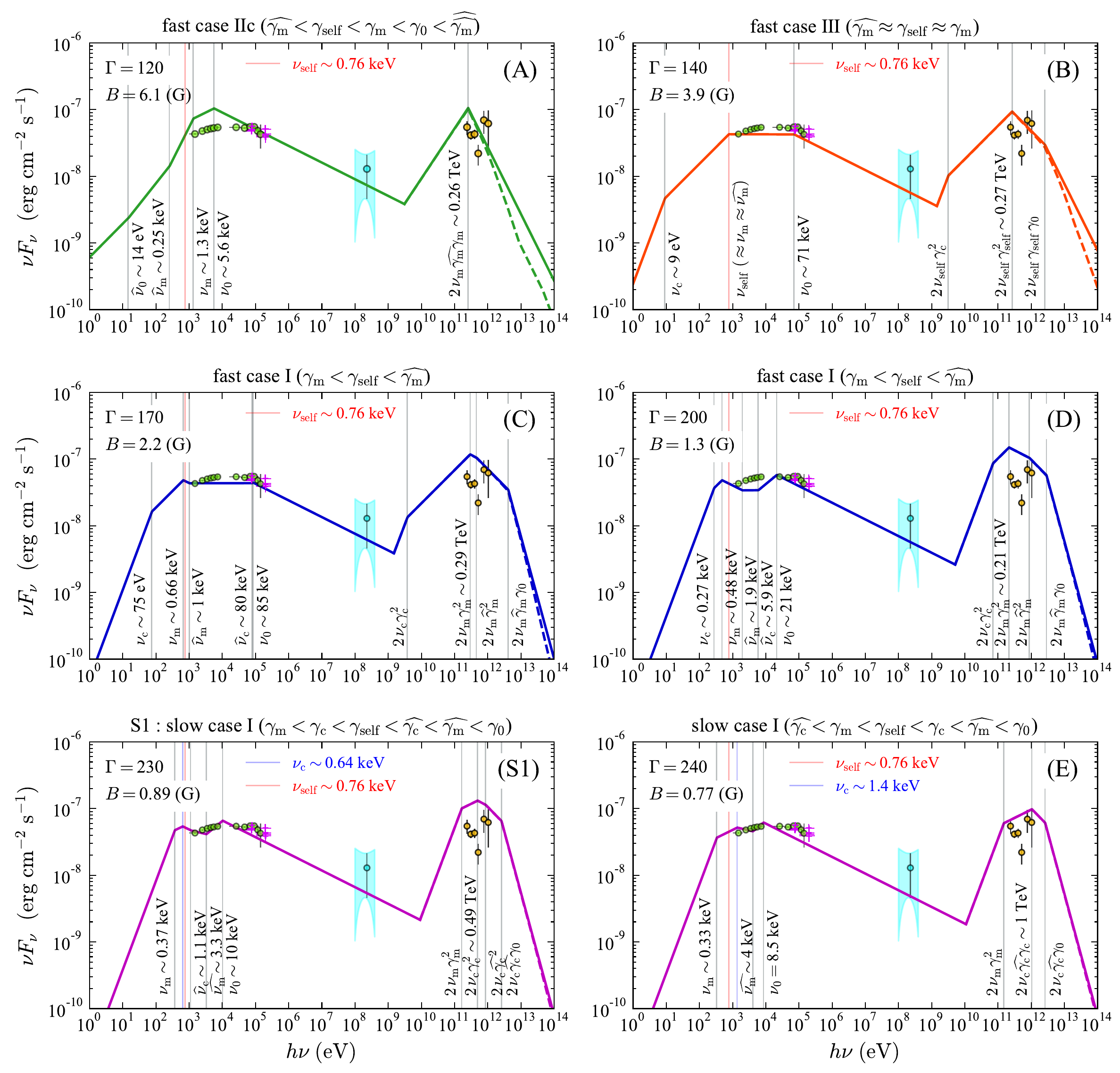}
 \caption{A sequence of early time model spectra corresponding to points along $B\propto\Gamma^{-3}$ track with $h\nu_{\rm self}\sim0.76$ keV shown in Figure \ref{fig:Gamma_B_map_summary} ($\epsilon_e/\epsilon_B=9.6$ and $\gm=1.5\times10^4$ are unchanged). A different color represents a different spectral regime. {The dashed lines show the effect of self absorption, which becomes less and less significant as $\Gamma$ increases.}}
\label{fig:spectra_mosaic}
\end{figure*}

\subsection{Solution Tracks on $\Gamma$--$B$ Plane}
\label{ss:solution track}

In \S \ref{ss:early SED}, we have derived analytic solutions with specific spectral cases both in fast (FIIc) and slow cooling (SI) regimes. However, this does not necessarily mean that there are no reasonable approximate solutions in other spectral regimes. 
Recently, DP21 numerically found that while the best fit solution is around our analytic solution,
there is a continuous sequence of numerical solutions in the $\Gamma$--$B$ plane (see the left panel of Figure \ref{fig:Gamma_B_map_summary}) for which one can obtain a reasonable  $\chi^2$ fit of the model to the X-rays and TeV observed in GRB 190114C. DP21 find that the this is broken by adding a constraint from the optical data point, we don't consider here. This constraint strongly favors the fast cooling solution.} This  numerical solution valley that gives  acceptable $\chi^2$ values on the $\Gamma$--$B$ plane roughly follows the scaling relation $B \propto \Gamma^{-3} $
while the remaining two parameters $\gm$ and $\epsilon_e/\epsilon_B$ do not change by a large factor. In the framework of analytic SSC model, this scaling relation is equivalent to the condition
\beqn
\nu_{\rm self}=\nu_{\rm syn}(\gamma_{\rm self})\propto B\gamma_{\rm self}^2\Gamma\propto B^{\frac{1}{3}}\Gamma = const\  ,
\eeqn
where $\nu_{\rm self}$ is the observed synchrotron frequency of electrons with KN-limit LF $\gamma_{\rm self}\propto B^{-1/3}$ (Eq. \eqref{eq:gamma_self}). 
Along this line of $B \propto \Gamma^{-3} $
the sum of the electron energy and the magnetic field energy $E_e+E_B = E_B(1+\epsilon_e/\epsilon_B)$ is conserved as long as $\epsilon_e/\epsilon_B$ is constant, since $E_B\propto B^2R^3\propto(B\Gamma^3)^2=const$, where $R\propto\Gamma^2$ is the shock radius.

The left panel of Figure \ref{fig:Gamma_B_map_summary} depicts the location of the analytic solutions to the early time data set (see \S \ref{ss:early SED}) as well as the numerical solution track (DP21) on the $\Gamma$--$B$ plane. Remarkably, the trajectory of $B\propto\Gamma^{-3}$ (or $\nu_{\rm self}=const$) that passes the analytic slow cooling solution S1 (see \S \ref{sss:slow}) is close to the numerical track, which implies that a continuous sequence of analytic acceptable solutions also exist along this line. 
Within the solutions presented the fast cooling solution marked A (see Figures \ref{fig:Gamma_B_map_summary} and \ref{fig:spectra_mosaic}) has the nearest flux to the optical point $\nu F_{\nu}(2 \ {\rm eV})\approx 10^{-9}\ {\rm erg\ cm^2\ s^{-1}}$. However, unlike DP21 this analytic model does not include the important contribution of the secondary pairs to the optical emission.  

Figure \ref{fig:spectra_mosaic} demonstrates the analytic spectra along the $B\propto\Gamma^{-3}$ track when fixing $\epsilon_e/\epsilon_B$ and $\gm$ to those for S1. As moving from the low-$\Gamma$ end (point A) toward the high-$\Gamma$ end (point D) along the track, the spectral regime undergoes transitions from the strong KN regime {FIIc} ($\gm/\hgm>1$) {through FIII ($\gm/\hgm=1$)} to the weak KN regime {FI} ($\gm/\hgm<1$) in the fast cooling since $\gm/\hgm\propto B\gm^3\propto \Gamma^{-3}$. Then, once $\Gamma$ becomes large enough, there is another transition from fast {(FI)} to slow {(SI)} cooling regimes{, followed by the slow strong KN regime SII}
(at some point between D and E) because of the rapid increase of $\gc$ while $\gm$ being constant. Apparently, all the spectra from A to E are marginally consistent with the observed early time spectrum. The low-$\Gamma$ end (near A) of the $B\propto\Gamma^{-3}$ trajectory falls into the same spectral type (FIIc) as the F2C. 
Solution A on ($\Gamma, B$) plane is located within the error range around F2C that we estimated in \S \ref{ss:uncertainty}.

To further investigate the origin of the similar spectral shapes along the  track, we show the evolution of break frequencies in the right panel of Figure \ref{fig:Gamma_B_map_summary}. While both the synchrotron peak frequency $\nu_{\rm s}$ and the IC peak frequency $\nu_{_{\rm IC}}$ stay almost constant at $\Gamma\lesssim200$, they start to increase monotonically as $\Gamma$ further increases in the SII regime, where $\nu_{\rm s}=\nu_{\rm c}\sim\nu_{\rm c}^{\rm syn}\propto B^{-3}\Gamma^{-1}\propto\Gamma^8$ and $\nu_{_{\rm IC}}=2\nu_{\rm c}\hgc\gc\propto\Gamma\gc\sim\Gamma\gamma_{\rm c}^{\rm syn}\propto B^{-2}\propto\Gamma^6$, becoming less and less consistent with the observation.

Figure \ref{fig:dp21} shows the analytic SSC spectrum using the parameters  of the best fit solution  obtained numerically by DP21 for \citep{sari98} dynamical parameters and ISM\footnote{These parameters are slightly different from those given in DP21, in which the best fit model depends on SEDs taken at two different times and the dynamical coefficients used are different.}: $\Gamma=110$, $B=6.5$ G, $\gm=19400$ and $\epsilon_e/\epsilon_B=31$. We mark this solution as {\it DP21}$_A$. 
Naturally, this analytic spectrum is different form the numerical one found by DP21 with the same parameters.  The analytic spectrum for {\it DP21}$_A$ has the same spectral type (FIIc) as F2C  and its location on ($\Gamma, B$) plane is within the error range around F2C. While the numerical solution of DP21 is, of course, consistent with the flux level,
the analytic solution, {\it DP21}$_A$, with the same parameters has a few times higher flux level compared to the data.   
Given the approximate nature of the analytic model we conclude that F2C obtained by an analytic approach is in reasonable agreement with the numerical solution of DP21.

\begin{figure}
\centering
\includegraphics[width=.5\textwidth]{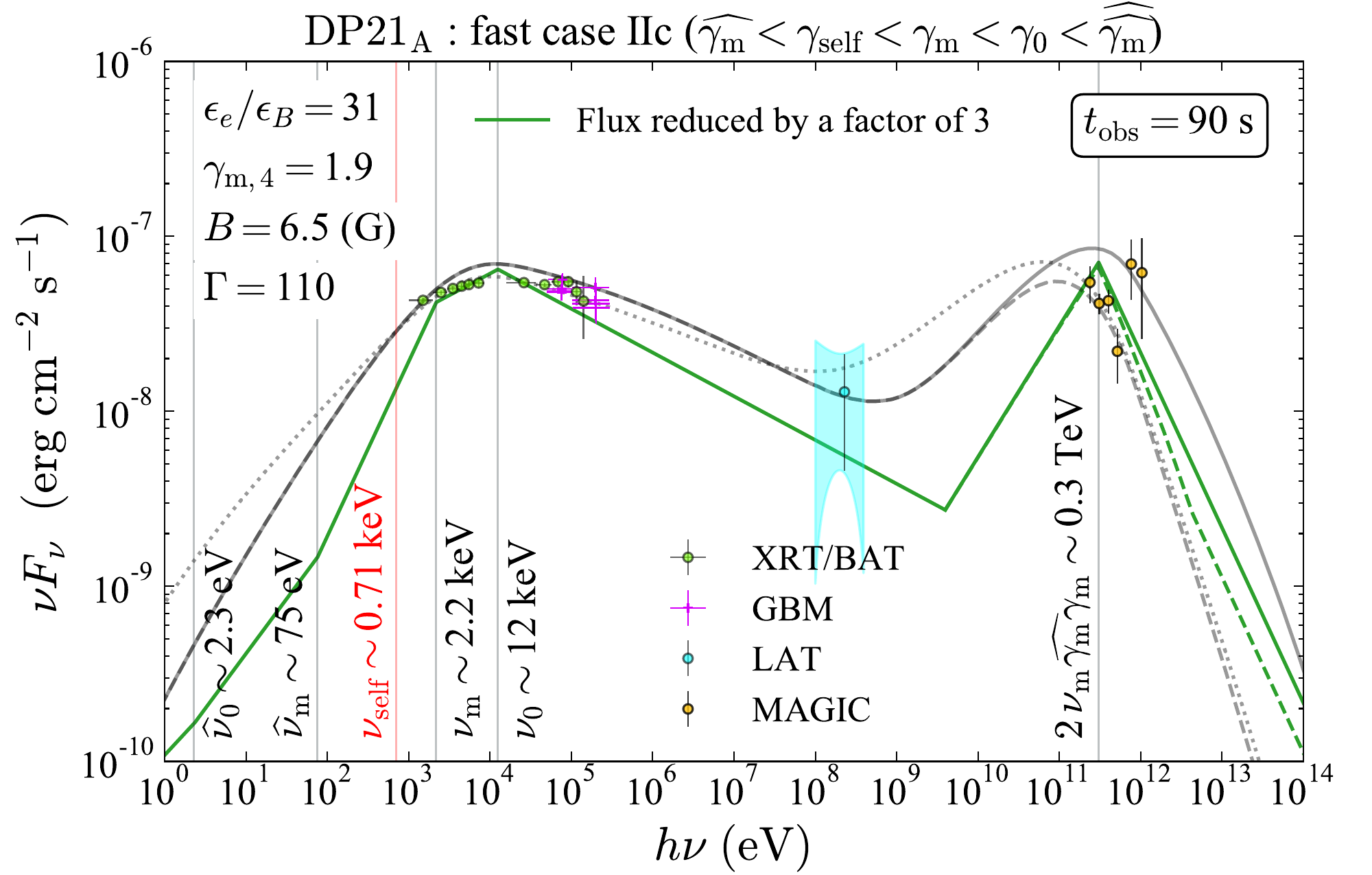}
 \caption{An analytic  spectrum using the best fit parameters obtained numerically by DP21 as compared to numerical spectra calculated by DP21. 
 The flux of the analytic spectrum has been  artificially reduced by a factor of three. 
 The grey curves represents the numerical spectra (E. Derishev, private communication). 
{The solid curves represent the analytic and numerical SEDs without self-absorption, whereas}
 the dashed curves depict the effect of self-absorption on the  SEDs.
 The dotted curve depicts the numerical SED that includes the effects of self-absorption as well as synchrotron and IC of the secondary pairs. The latter emission increases the low energy (eV) as well as the GeV fluxes (see Fig. 5 of DP21). }
\label{fig:dp21}
\end{figure}

\subsection{Solution with $f_{_{\rm KN}}=1$}
\label{ss:f_KN_1_solution}

Although $f_{_{\rm KN}}=0.2$ is adopted in our main analysis, below we briefly the solutions with the  conventional choice of $f_{_{\rm KN}}=1$ and  how the inferred model parameters vary. In particular, we consider a fast cooling solution for the early time spectrum at $t_{\rm obs}=90$ s. The characteristic fast-cooling solution with $f_{_{\rm KN}}=0.2$ for the early time data lies in the FIIc spectral regime, where the KN effect is relatively strong: $\gm/\hgm>(\epsilon_e/\epsilon_B)^{1/3}>1$ (see \S \ref{ss:early SED}). Since $\hgm\propto f_{_{\rm KN}}$, an increase of $f_{_{\rm KN}}$ from $0.2$ to $1.0$ decreases the ratio $\gm/\hgm$ by a factor of five, which would effectively shift the system toward the weaker KN regime. Consequently, we identify the relevant spectral case as the relatively weak KN regime FI defined by $\gm/\hgm<1$. 

In FI with $p=2.5$, the synchrotron spectrum peaks at $\nu_{\rm m}$ but with extended plateau between $\hnm$ and $\nu_0$, which makes the spectrum rather broad. To find an analytic solution in FI, we repeat the same procedure as done in \S \ref{ss:early SED} (Eqs. \eqref{eq:CaseIIc-eq1}--\eqref{eq:CaseIIc-eq4}) but with some modifications that hold in this regime: $\nu_{\rm s}=\nu_{\rm m}$, $\nuic=2\gamma_{\rm m}^2\,\nu_{\rm m}$, $Y_{\rm obs}=Y(\gm)$, and $F_{\nu_{\rm s}}=F_{\nu_{\rm m}}$. Here we show the result with an approximation $Y(\gm)\approx[(\epsilon_e/\epsilon_B)\,(p-2)]^{1/2}\,(\gm/\hgm)^{-1/2}$, which is valid for $p-2<\gm/\hgm<(\epsilon_e/\epsilon_B)^2$. Although we adopt this case as we are interested in solutions in which $\gm$ , $\gamma_{\rm self}$ and $\hgm$ are relatively close to each other, one can also try a simpler approximation $Y(\gm)\approx(\epsilon_e/\epsilon_B)^{1/2}$, which is valid for $\gm/\hgm<p-2$, to get almost the same result as we present below (see Eq. (42) in NAS09 for the detailed approximation of $Y(\gm)$).
We obtain a solution by choosing observational parameters such that $h\nu_{\rm s}=2.0$ keV, $h\nuic=0.1$ TeV, $Y_{\rm obs}=2.0$ and $F_{\nu_{\rm s}}=12$ mJy:
\beqn
\frac{\epsilon_e}{\epsilon_B}&=&(p-2)^{-\frac{8}{7}}\,C_1^{-\frac{12}{7}}\,C_2^{-\frac{8}{7}}\,C_3^{\frac{2}{7}}\,\nu_{\rm s}^{\frac{10}{7}}\,F_{\nu_{\rm s}}^{-\frac{2}{7}}\,Y_{\rm obs}^{\frac{16}{7}}\nonumber\\&=&6.1\ \left(\frac{f_{_{\rm KN}}}{1.0}\right)^{-\frac{8}{7}}\left(\frac{h\nu_{\rm s}}{2.0{\, \rm keV}}\right)^{\frac{10}{7}}\left(\frac{Y_{\rm obs}}{2.0}\right)^{\frac{16}{7}}\left(\frac{F_{\nu_{\rm s}}}{12{\, \rm mJy}}\right)^{-\frac{2}{7}},
\eeqn
\vspace{-0.3cm}
\beqn
\gm &=&2^{-\frac{1}{2}}\,\nu_{\rm s}^{-\frac{1}{2}}\,\nu_{_{\rm IC}}^{\frac{1}{2}}\nonumber\\&=& 5.0\times10^3\ \left(\frac{h\nu_{\rm s}}{2.0{\, \rm keV}}\right)^{-\frac{1}{2}}\left(\frac{h\nu_{_{\rm IC}}}{0.1{\, \rm TeV}}\right)^{\frac{1}{2}},
\eeqn
\vspace{-0.3cm}
\beqn
B &=&2^{\frac{3}{2}}\,(p-2)^{-\frac{1}{7}}\,C_1^{-\frac{12}{7}}\,C_2^{-\frac{1}{7}}\,C_3^{\frac{2}{7}}\,\nu_{\rm s}^{\frac{41}{14}}\,\nu_{_{\rm IC}}^{-\frac{3}{2}}\,F_{\nu_{\rm s}}^{-\frac{2}{7}}\,Y_{\rm obs}^{\frac{2}{7}}\nonumber\\&=& 270\ {\rm G}\ \left(\frac{f_{_{\rm KN}}}{1.0}\right)^{-\frac{1}{7}}\left(\frac{h\nu_{\rm s}}{2.0{\, \rm keV}}\right)^{\frac{41}{14}}\left(\frac{h\nu_{_{\rm IC}}}{0.1{\, \rm TeV}}\right)^{-\frac{3}{2}}\\&&\times\left(\frac{Y_{\rm obs}}{2.0}\right)^{\frac{2}{7}}\left(\frac{F_{\nu_{\rm s}}}{12{\, \rm mJy}}\right)^{-\frac{2}{7}},\nonumber
\eeqn
\vspace{-0.3cm}
\beqn
\Gamma &=&2^{-\frac{1}{2}}\,(p-2)^{\frac{1}{7}}\,C_1^{\frac{5}{7}}\,C_2^{\frac{1}{7}}\,C_3^{-\frac{2}{7}}\,\nu_{\rm s}^{-\frac{13}{14}}\,\nu_{_{\rm IC}}^{\frac{1}{2}}\,F_{\nu_{\rm s}}^{\frac{2}{7}}\,Y_{\rm obs}^{-\frac{2}{7}}\nonumber\\&=& 37\ \left(\frac{f_{_{\rm KN}}}{1.0}\right)^{\frac{1}{7}}\left(\frac{h\nu_{\rm s}}{2.0{\, \rm keV}}\right)^{-\frac{13}{14}}\left(\frac{h\nu_{_{\rm IC}}}{0.1{\, \rm TeV}}\right)^{\frac{1}{2}}\\&&\times\left(\frac{Y_{\rm obs}}{2.0}\right)^{-\frac{2}{7}}\left(\frac{F_{\nu_{\rm s}}}{12{\, \rm mJy}}\right)^{\frac{2}{7}},\nonumber
\eeqn
where $p=2.5$ 
have been used in the 
second equality of each equation. 
Figure \ref{fig:YP21fast_fKN1} shows the synchrotron-SSC spectrum corresponding to this solution. Compared to the parameters inferred from the former analysis with our fiducial choice $f_{_{\rm KN}}=0.2$ (as shown in Eqs. \eqref{eq:CaseIIc-eps}--\eqref{eq:CaseIIc-Gamma}), 
the solution with $f_{_{\rm KN}}=1$ results in a much larger $B$ and a smaller $\Gamma$ than those for $f_{_{\rm KN}}=0.2$. While the spectral fit looks reasonable, such a low $\Gamma$ value is inconsistent with the self absorption constraint leading to a severe suppression of IC component due to the internal absorption.

\begin{figure}
\centering
\includegraphics[width=.5\textwidth]{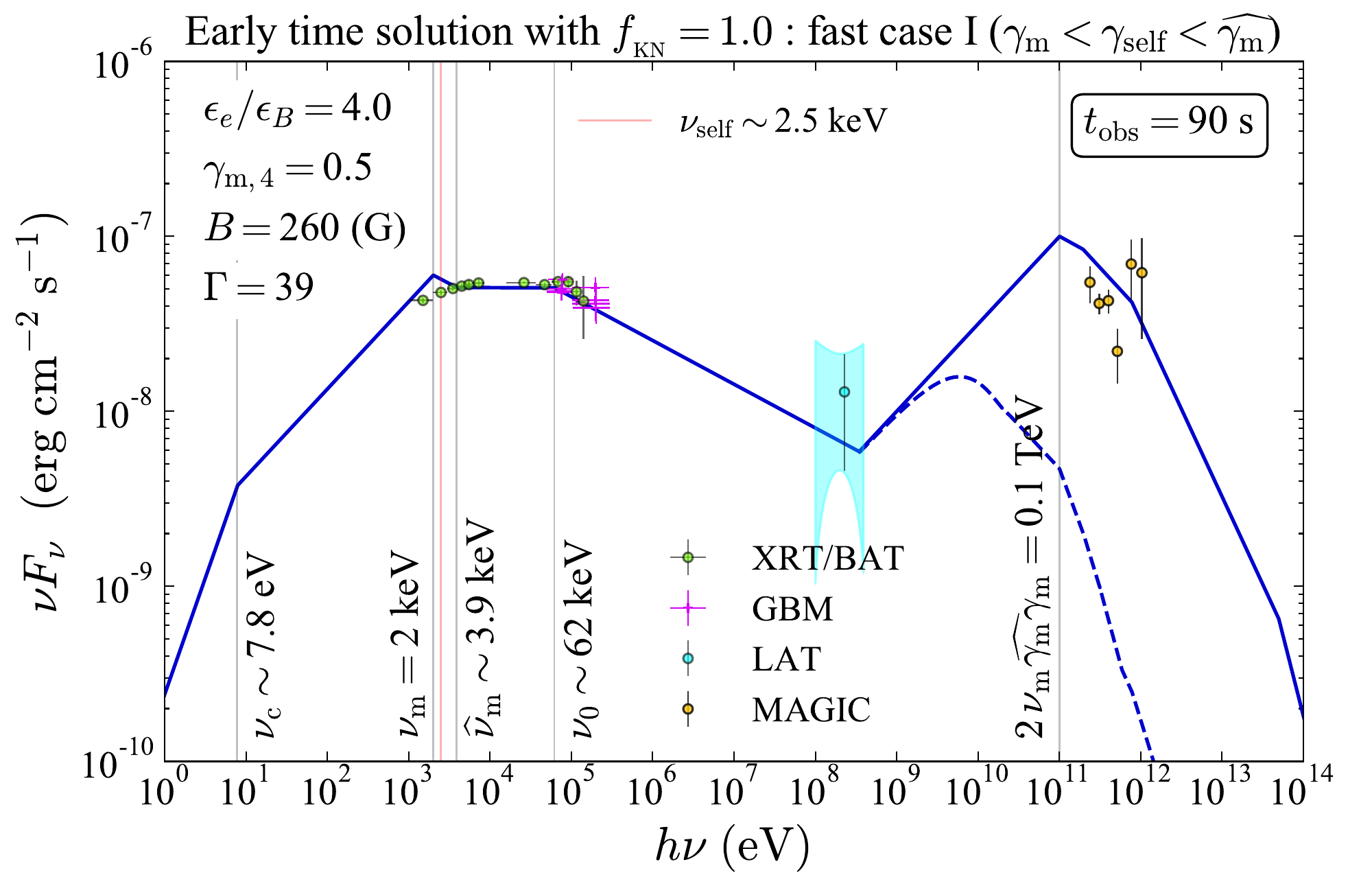}
\caption{Same as Figure \ref{fig:YP21fast} but with a conventional choice of $f_{_{\rm KN}}=1.0$. The model spectrum is for the analytic fast-cooling solution obtained by requiring $h\nu_{\rm s}=2.0$ keV, $h\nu_{_{\rm IC}}=0.1$ TeV, $Y_{\rm obs}=2.0$, $F_{\nu_{\rm s}}=12.4$ mJy with $p=2.5$. {The IC spectrum with self absorption is also shown in dashed line.}
While the analytic spectrum looks consistent with the data the obtained parameters are very different from those found in the numerical solution and are physically unlikely as with such a low $\Gamma$ self absorption of the IC photons would be too strong.  
}
\label{fig:YP21fast_fKN1}
\end{figure}

\section{Comparison with other work}
\label{s:comparison}

It is interesting to compare the analytic approach to numerical ones \citep{magic19,wang19,asano20}. To this end we try to analytically reproduce some of the numerically obtained solutions to the GRB 190114C afterglow spectra \citep{magic19,asano20}, and compare them with the analytic spectra for the same parameters. 

The model parameters given by \citet{asano20} are: an ISM density $n_0=0.3\ {\rm cm^{-3}}$, a total kinetic energy (isotropic equivalent)  $E_{0}=4\times10^{53}$ erg, $\epsilon_e=0.1$, 
$\epsilon_B=1\times10^{-3}$, $p=2.3$, and $t_{\rm obs}=100$ s. With these values we calculated the corresponding model parameters: $\Gamma=160$, $B=1.1$ G, $\gm=6900$ and $\epsilon_e/\epsilon_B=100$. The corresponding  solution seems to be  in the FI spectral regime. However,  \citet{asano20} assumes that the escape time of photons is a factor of $6$ times shorter than the lifetime of electrons. 
When modifying the analytic model to take this into account we find that the spectral regime shifts from FI to SI. 
Figure \ref{fig:Asano20} shows the analytically-reproduced spectrum (``Asano20'') with the same model parameters as those used in the numerical model in comparison with the numerical model and the observation.  The analytic spectrum shows a saddle-shaped decrease in the flux of the synchrotron component by a factor of about two at frequencies between $\nu_{\rm c}$ and $\nu_0$ due to the KN effect, making the modeled synchrotron component seemingly inconsistent with the observation. However, if the approximate spectrum is smoothed around the break frequencies, which is the case for true spectra taking the actual KN cross section into account, then the shape of the synchrotron peak may be relatively flat at $\nu_{\rm c}<\nu<\nu_0$ and consistent with the observed one.  
Thus, the analytically-reproduced spectra is at least in qualitative agreement with observation.
Also in this case the analytic spectrum with $f_{_{\rm KN}}=0.2$ is in much better agreement with  the numerical model than the spectrum with $f_{_{\rm KN}}=1$. 

{The model parameters given by \citet{magic19} are: $n_0=0.5\ {\rm cm^{-3}}$, $E_{0}=8\times10^{53}$ erg, $\epsilon_e=0.07$, 
$\epsilon_B=8\times10^{-5}$, $p=2.6$, and $t_{\rm obs}=90$ s (in the middle of the quoted time
interval) and  the corresponding model parameters are: $\Gamma=170$, $B=0.43$ G, $\gm=8400$ and $\epsilon_e/\epsilon_B=880$. 
Figure \ref{fig:magic19} shows the analytically-reproduced spectrum (``MAGIC19'') in comparison with the numerical model and the observation. The analytic solution  MAGIC19 falls into the SI spectral regime. However,  it is evidently inconsistent with both the observation and the numerical model in terms of the value of $Y_{\rm obs}$ and the shape of the overall spectrum.  The extremely large  $\epsilon_e/\epsilon_B=880$ value reported by \citet{magic19} is most likely the source of the huge discrepancy in $Y_{\rm obs}$ by a factor of $25$.
A similar disagreement is also confirmed by other numerical works (\citealt{asano20}; DP21). Thus, we conclude that it is not due to the approximated nature of the analytic SSC model.

\begin{figure}
\centering
\includegraphics[width=.5\textwidth]{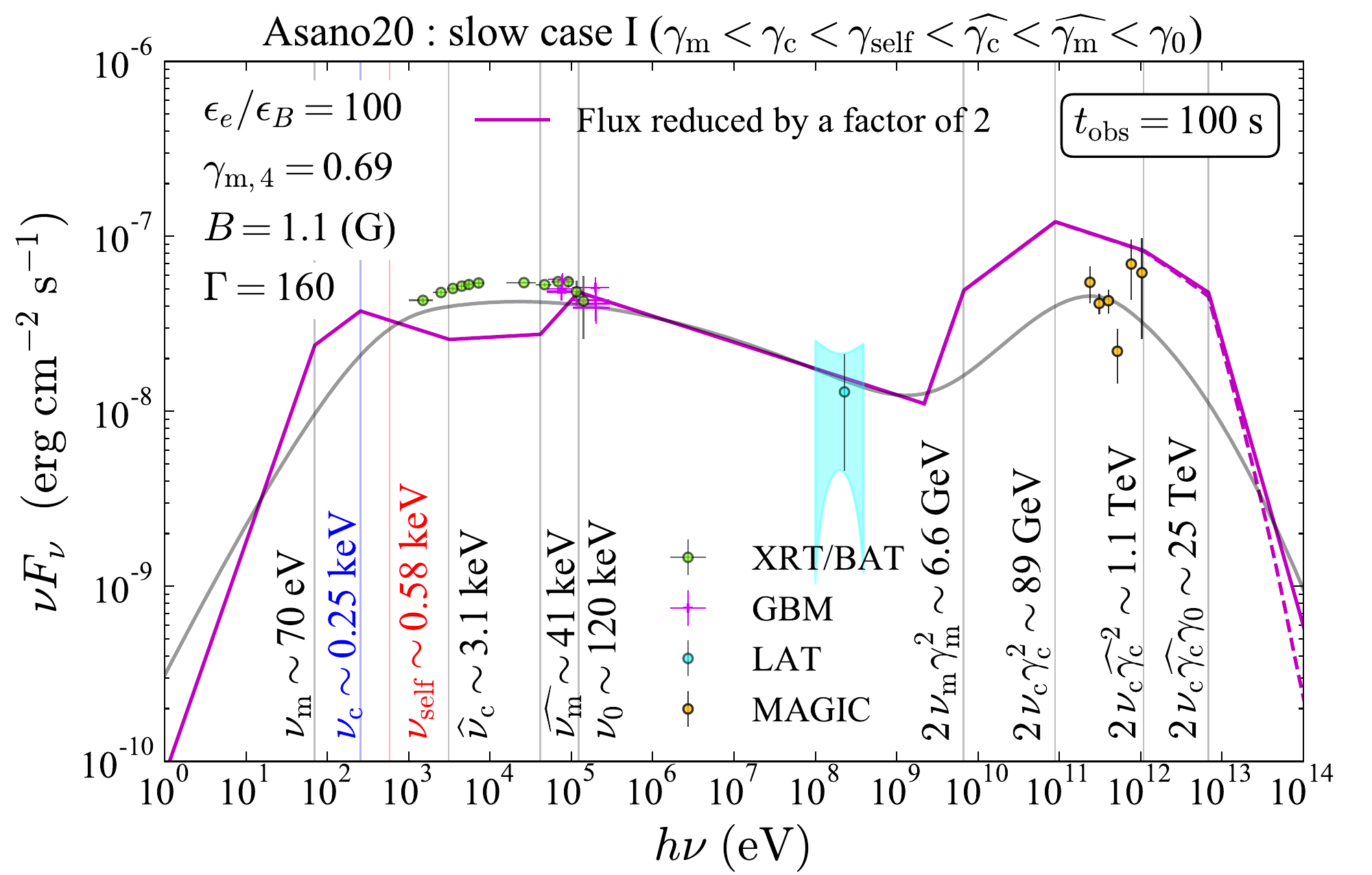}
 \caption{An analytic reproduction of SSC spectrum numerically obtained by \citet{asano20} (the model with a time-independent electron spectrum called ``ISM (method II)'' in their paper based on \citealt{zhang20,murase11}) with some modifications (see the main text) with the flux artificially reduced by a factor of 2. The IC spectrum with self absorption is also shown in dashed line. {The grey curve represents the numerical spectrum (K. Murase, private communication).}
 }
\label{fig:Asano20}
\end{figure}

\begin{figure}
\centering
\includegraphics[width=.5\textwidth]{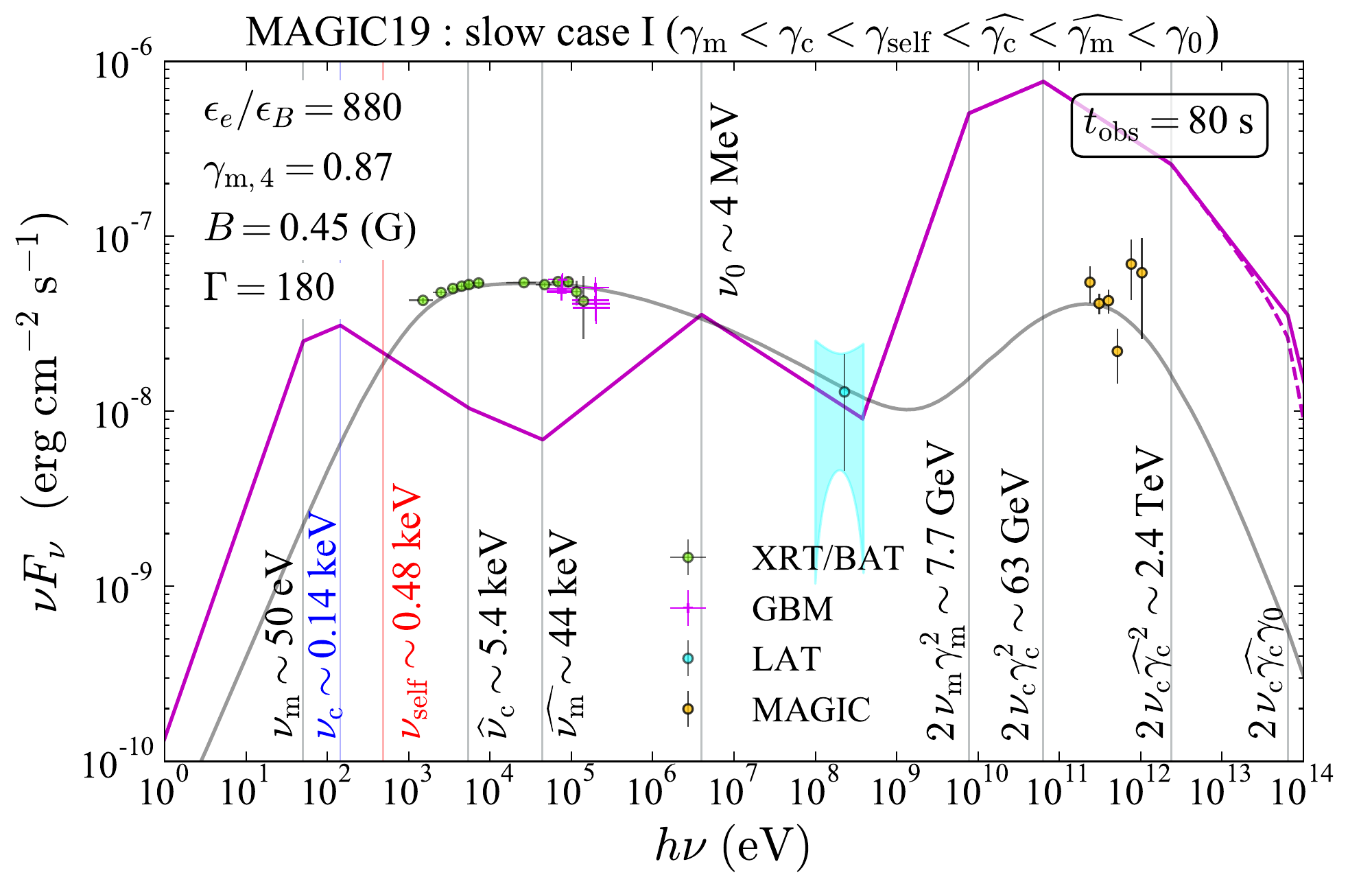}
 \caption{An analytic reproduction of SSC spectrum numerically obtained by \citet{magic19}. 
 The grey curve represents the model spectrum produced by \citet{magic19}. {The IC spectrum with self absorption is also shown in dashed line.}
 }
 \label{fig:magic19}
\end{figure}

\section{Summary and Discussion}
\label{s:discussion}

In this work, we estimate the physical conditions of the emitting region in a GRB afterglow from which both low energy (synchrotron) and high energy (IC) emission are observed.  We apply the analytic SSC model of NAS09 that includes the KN effects  to the multi-wavelength (X-rays and TeV gamma-rays) afterglow spectrum of GRB 190114C. 
Following DP21, we use four model parameters to describe the emitting region: $\gm$, $\Gamma$, $B$, and $\epsilon_e/\epsilon_B$ (with $p=2.5$). We stress that the common assumption $\gm\propto\Gamma$ has not been made in our modelling.  We also don't assume that the number of emitting  (accelerated) electrons equals the number of electrons that the blast waves has engulfed. We also stress that with minimal appropriate modifications the method can be applied to other astronomical systems. 

Our findings are summarized as follows:

\begin{itemize}
    \item  
    We introduce a modification of the common approximation concerning the KN regime, $f_{_{\rm KN}}=0.2$, which turns out to be essential to obtain a better consistency between analytic and numerical parameters. Importantly one can obtain reasonable solutions with $f_{_{\rm KN}}=1$. However, the obtained solution parameters are unreasonable. In particular  $\Gamma=39$ which  would lead to a very significant absorption of the TeV signal.
    \item 
    We find the characteristic solutions for the GRB 190114C afterglow SED at $t_{\rm obs}=90$ s in both fast-cooling regime with strong KN effects and slow-cooling regime with weak KN effects. Considering the quality of the fit to the SSC component to the data, we conclude that the fast cooling solution in relatively strong KN regime (FIIc) is favored over the slow cooling solution for the early time observations. These results are consistent with the numerical solution of DP21.
    
    \item 
    We also confirm the numerical findings of DP21 that along a  track between these two solutions at $100 \le \Gamma \le 250$  there is a sequence of solutions that are marginally consistent with the observed spectrum of GRB 190114C. Such a track arises naturally within the fast cooling regime (FIIc), in which the best fit solution is located.

    \item 
    Once $f_{_{\rm KN}} $ is introduced, the analytically-reproduced spectra for parameters inferred from numerical solutions by \citet{asano20} and \citet{dp21} are in  qualitatively good agreement (up to factors of $2$--$3$ in the resulting flux) with the observations. Nevertheless, the analytic solution is inconsistent with those  presented in the discovery paper \cite{magic19} (see also DP21).
    
    \item  Our motivation was to establish a methodology for determining the analytic solutions for a given afterglow data set consisting of X-ray and TeV components.  However, despite the relative convenience of the analytic SSC model, we find that the analytic SSC modeling is burdened by generic problems. First, one has to select the  spectral regime before deriving the solutions. The solution obtained could be inconsistent with the regime assumed. Second,  the solution of the analytical model, especially $\epsilon_e/\epsilon_B$ which plays a major role in determining  the spectral regime, is highly sensitive to the observed parameters. This leads to a technical complication, adopting slightly different observed parameters even within the uncertainty of the observation may move the solution from  one spectral regime to another. Moreover, we find that the inferred parameters, and even the existence of a solution within a given spectral range, depend strongly on the  quantities adopted to characterize the observed spectrum.  We have shown in \S \ref{ss:uncertainty} that the obtained parameters can vary significantly  while varying the observed characteristics of the spectrum within their  the error range. This last point is critical as given the piece-wise linear (in the log-log plane) nature of the analytic solution we are forced to select from the data typical peak frequencies and fluxes to characterize the observations.
    
\end{itemize}

Finally, potential caveats to our method are as follows. First, a limitation of the simple analytic approach in determining the afterglow parameters is that it cannot take into account for the effects of self-absorption  within the source.
{While this can be done with minor numerical modifications, the emission arising from the resulting pairs can be significant (DP21) and this requires detailed numerical simulations. }
This can significantly alter the spectrum. 
{For GRB 190114C this is mostly important in the eV and the GeV regimes.}

In conclusion, while the analytical SSC model suffers from some general problems in determining the parameters of afterglow, these approximations provide a useful tool for interpreting the observations without the need of elaborated numerical calculations. 

\section*{Acknowledgements}
We thank Katsuaki Asano, Evgeney Derishev, Julian Krolik and Kohta Murase for useful discussion and for sharing their model data. We also thank the anonymous referee for carefully reading our manuscript. This work is supported by the advanced ERC grant TReX.

\section*{Data Availability}
This is a theoretical paper that does not involve any new data. The model data presented in this article are all reproducible.

\bibliographystyle{mnras}

\begin{thebibliography}{}
\makeatletter
\relax
\def\mn@urlcharsother{\let\do\@makeother \do\$\do\&\do\#\do\^\do\_\do\%\do\~}
\def\mn@doi{\begingroup\mn@urlcharsother \@ifnextchar [ {\mn@doi@}
  {\mn@doi@[]}}
\def\mn@doi@[#1]#2{\def\@tempa{#1}\ifx\@tempa\@empty \href
  {http://dx.doi.org/#2} {doi:#2}\else \href {http://dx.doi.org/#2} {#1}\fi
  \endgroup}
\def\mn@eprint#1#2{\mn@eprint@#1:#2::\@nil}
\def\mn@eprint@arXiv#1{\href {http://arxiv.org/abs/#1} {{\tt arXiv:#1}}}
\def\mn@eprint@dblp#1{\href {http://dblp.uni-trier.de/rec/bibtex/#1.xml}
  {dblp:#1}}
\def\mn@eprint@#1:#2:#3:#4\@nil{\def\@tempa {#1}\def\@tempb {#2}\def\@tempc
  {#3}\ifx \@tempc \@empty \let \@tempc \@tempb \let \@tempb \@tempa \fi \ifx
  \@tempb \@empty \def\@tempb {arXiv}\fi \@ifundefined
  {mn@eprint@\@tempb}{\@tempb:\@tempc}{\expandafter \expandafter \csname
  mn@eprint@\@tempb\endcsname \expandafter{\@tempc}}}

\bibitem[\protect\citeauthoryear{{Aharonian}}{{Aharonian}}{2000}]{Aharonian2000}
{Aharonian} F.~A.,  2000, \mn@doi [\na] {10.1016/S1384-1076(00)00039-7}, \href
  {https://ui.adsabs.harvard.edu/abs/2000NewA....5..377A} {5, 377}

\bibitem[\protect\citeauthoryear{{Ajello} et~al.,}{{Ajello}
  et~al.}{2020}]{ajello20}
{Ajello} M.,  et~al., 2020, \mn@doi [\apj] {10.3847/1538-4357/ab5b05}, \href
  {https://ui.adsabs.harvard.edu/abs/2020ApJ...890....9A} {890, 9}

\bibitem[\protect\citeauthoryear{{Ando}, {Nakar}  \& {Sari}}{{Ando}
  et~al.}{2008}]{ando08}
{Ando} S.,  {Nakar} E.,   {Sari} R.,  2008, \mn@doi [\apj] {10.1086/592486},
  \href {https://ui.adsabs.harvard.edu/abs/2008ApJ...689.1150A} {689, 1150}

\bibitem[\protect\citeauthoryear{{Asano}, {Murase}  \& {Toma}}{{Asano}
  et~al.}{2020}]{asano20}
{Asano} K.,  {Murase} K.,   {Toma} K.,  2020, arXiv e-prints, \href
  {https://ui.adsabs.harvard.edu/abs/2020arXiv200706307A} {p. arXiv:2007.06307}

\bibitem[\protect\citeauthoryear{{De Jager}, {Harding}, {Michelson}, {Nel},
  {Nolan}, {Sreekumar}  \& {Thompson}}{{De Jager} et~al.}{1996}]{deJager1996}
{De Jager} O.~C.,  {Harding} A.~K.,  {Michelson} P.~F.,  {Nel} H.~I.,  {Nolan}
  P.~L.,  {Sreekumar} P.,   {Thompson} D.~J.,  1996, \mn@doi [\apj]
  {10.1086/176726}, \href
  {https://ui.adsabs.harvard.edu/abs/1996ApJ...457..253D} {457, 253}

\bibitem[\protect\citeauthoryear{{Derishev} \& {Piran}}{{Derishev} \&
  {Piran}}{2019}]{dp19}
{Derishev} E.,  {Piran} T.,  2019, \mn@doi [\apjl] {10.3847/2041-8213/ab2d8a},
  \href {https://ui.adsabs.harvard.edu/abs/2019ApJ...880L..27D} {880, L27}

\bibitem[\protect\citeauthoryear{{Derishev} \& {Piran}}{{Derishev} \&
  {Piran}}{2021}]{dp21}
{Derishev} E.,  {Piran} T.,  2021, arXiv e-prints, \href
  {https://ui.adsabs.harvard.edu/abs/2021arXiv210612035D} {p. arXiv:2106.12035}

\bibitem[\protect\citeauthoryear{{Derishev}, {Kocharovsky}  \&
  {Kocharovsky}}{{Derishev} et~al.}{2001}]{Derishev2001}
{Derishev} E.~V.,  {Kocharovsky} V.~V.,   {Kocharovsky} V.~V.,  2001, \mn@doi
  [\aap] {10.1051/0004-6361:20010586}, \href
  {https://ui.adsabs.harvard.edu/abs/2001A&A...372.1071D} {372, 1071}

\bibitem[\protect\citeauthoryear{{Derishev}, {Kocharovsky}, {Kocharovsky}  \&
  {M{\'e}sz{\'a}ros}}{{Derishev} et~al.}{2003}]{Derishev2003}
{Derishev} E.~V.,  {Kocharovsky} V.~V.,  {Kocharovsky} V.~V.,
  {M{\'e}sz{\'a}ros} P.,  2003, in {Ricker} G.~R.,  {Vanderspek} R.~K.,  eds,
  American Institute of Physics Conference Series Vol. 662, Gamma-Ray Burst and
  Afterglow Astronomy 2001: A Workshop Celebrating the First Year of the HETE
  Mission. pp 292--294, \mn@doi{10.1063/1.1579362}

\bibitem[\protect\citeauthoryear{{Guilbert}, {Fabian}  \& {Rees}}{{Guilbert}
  et~al.}{1983}]{Guilbert1983}
{Guilbert} P.~W.,  {Fabian} A.~C.,   {Rees} M.~J.,  1983, \mn@doi [\mnras]
  {10.1093/mnras/205.3.593}, \href
  {https://ui.adsabs.harvard.edu/abs/1983MNRAS.205..593G} {205, 593}

\bibitem[\protect\citeauthoryear{{H.~E.~S.~S. Collaboration}
  et~al.,}{{H.~E.~S.~S. Collaboration} et~al.}{2021}]{hess21}
{H.~E.~S.~S. Collaboration} et~al., 2021, \mn@doi [Science]
  {10.1126/science.abe8560}, \href
  {https://ui.adsabs.harvard.edu/abs/2021Sci...372.1081H} {372, 1081}

\bibitem[\protect\citeauthoryear{{MAGIC Collaboration} et~al.,}{{MAGIC
  Collaboration} et~al.}{2019}]{magic19}
{MAGIC Collaboration} et~al., 2019, \mn@doi [\nat] {10.1038/s41586-019-1754-6},
  \href {https://ui.adsabs.harvard.edu/abs/2019Natur.575..459M} {575, 459}

\bibitem[\protect\citeauthoryear{{Murase}, {Toma}, {Yamazaki}  \&
  {M{\'e}sz{\'a}ros}}{{Murase} et~al.}{2011}]{murase11}
{Murase} K.,  {Toma} K.,  {Yamazaki} R.,   {M{\'e}sz{\'a}ros} P.,  2011,
  \mn@doi [\apj] {10.1088/0004-637X/732/2/77}, \href
  {https://ui.adsabs.harvard.edu/abs/2011ApJ...732...77M} {732, 77}

\bibitem[\protect\citeauthoryear{{Nakar}, {Ando}  \& {Sari}}{{Nakar}
  et~al.}{2009}]{nakar09}
{Nakar} E.,  {Ando} S.,   {Sari} R.,  2009, \mn@doi [\apj]
  {10.1088/0004-637X/703/1/675}, \href
  {https://ui.adsabs.harvard.edu/abs/2009ApJ...703..675N} {703, 675}

\bibitem[\protect\citeauthoryear{{Sari} \& {Esin}}{{Sari} \&
  {Esin}}{2001}]{sari01}
{Sari} R.,  {Esin} A.~A.,  2001, \mn@doi [\apj] {10.1086/319003}, \href
  {https://ui.adsabs.harvard.edu/abs/2001ApJ...548..787S} {548, 787}

\bibitem[\protect\citeauthoryear{{Sari}, {Narayan}  \& {Piran}}{{Sari}
  et~al.}{1996}]{snp96}
{Sari} R.,  {Narayan} R.,   {Piran} T.,  1996, \mn@doi [\apj] {10.1086/178136},
  \href {https://ui.adsabs.harvard.edu/abs/1996ApJ...473..204S} {473, 204}

\bibitem[\protect\citeauthoryear{{Sari}, {Piran}  \& {Narayan}}{{Sari}
  et~al.}{1998}]{sari98}
{Sari} R.,  {Piran} T.,   {Narayan} R.,  1998, \mn@doi [\apjl]
  {10.1086/311269}, \href
  {https://ui.adsabs.harvard.edu/abs/1998ApJ...497L..17S} {497, L17}

\bibitem[\protect\citeauthoryear{{Wang}, {Liu}, {Zhang}, {Xi}  \&
  {Zhang}}{{Wang} et~al.}{2019}]{wang19}
{Wang} X.-Y.,  {Liu} R.-Y.,  {Zhang} H.-M.,  {Xi} S.-Q.,   {Zhang} B.,  2019,
  \mn@doi [\apj] {10.3847/1538-4357/ab426c}, \href
  {https://ui.adsabs.harvard.edu/abs/2019ApJ...884..117W} {884, 117}

\bibitem[\protect\citeauthoryear{{Zhang}, {Murase}, {Veres}  \&
  {M{\'e}sz{\'a}ros}}{{Zhang} et~al.}{2021}]{zhang20}
{Zhang} B.~T.,  {Murase} K.,  {Veres} P.,   {M{\'e}sz{\'a}ros} P.,  2021,
  \mn@doi [\apj] {10.3847/1538-4357/ac0cfc}, \href
  {https://ui.adsabs.harvard.edu/abs/2021ApJ...920...55Z} {920, 55}

\makeatother
\end{thebibliography}
\input{analyticSSC.bbl}

\bsp	
\label{lastpage}
\end{document}